\documentclass[twocolumn,english,twocolumn,prl]{revtex4-1}
\usepackage[T1]{fontenc}
\usepackage[latin9]{inputenc}
\usepackage{geometry}
\usepackage{verbatim}
\usepackage{mathtools}
\usepackage{amsmath}
\usepackage{amssymb}
\usepackage{graphicx}

\makeatletter
\usepackage{amsmath}
\usepackage{subfigure}
\usepackage{graphicx, mathtools}
\usepackage[normalem]{ulem}
\usepackage[colorlinks=true,linkcolor=MidnightBlue,urlcolor=MidnightBlue,citecolor=MidnightBlue,anchorcolor=MidnightBlue]{hyperref}\usepackage[dvipsnames]{xcolor}

\usepackage{chngcntr}
\usepackage{listings}
\usepackage{xcolor, tikz-cd}
 
\definecolor{codegreen}{rgb}{0,0.6,0}
\definecolor{codegray}{rgb}{0.5,0.5,0.5}
\definecolor{codepurple}{rgb}{0.58,0,0.82}
\definecolor{backcolour}{rgb}{0.98,0.98,0.98}
 
\lstdefinestyle{mystyle}{
    backgroundcolor=\color{backcolour},   
    commentstyle=\color{codegreen},
    keywordstyle=\color{magenta},
    numberstyle=\tiny\color{codegray},
    stringstyle=\color{codepurple},
    basicstyle=\linespread{0.8}\footnotesize\ttfamily
    breakatwhitespace=false,         
    breaklines=true,                 
    captionpos=b,                    
    keepspaces=true,                 
    numbersep=5pt,                  
    showspaces=false,                
    showstringspaces=false,
    showtabs=false,                  
    tabsize=2
}

\makeatother

\usepackage{babel}
\begin{document}
\title{Diffusivity dependence of the transition path ensemble}
\author{Lukas Kikuchi}
\email{ltk26@cam.ac.uk }

\affiliation{DAMTP, Centre for Mathematical Sciences, University of Cambridge,
Wilberforce Road, Cambridge CB3 0WA, UK}
\author{Ronojoy Adhikari}
\affiliation{DAMTP, Centre for Mathematical Sciences, University of Cambridge,
Wilberforce Road, Cambridge CB3 0WA, UK}
\author{Julian Kappler}

\affiliation{DAMTP, Centre for Mathematical Sciences, University of Cambridge,
Wilberforce Road, Cambridge CB3 0WA, UK}
\begin{abstract}
	At low temperatures transition pathways of stochastic dynamical systems
	are typically approximated by instantons. Here we show, using a dynamical
	system containing two competing pathways, that even at low-to-intermediate
	temperatures, instantons can fail to capture the most likely transition
	pathway. We consider an approximation which includes Gaussian fluctuations
	around instantons and, by comparing with the results of an accurate
	and efficient path-space Monte Carlo sampling method, find this approximation
	to hold for a wide range of temperatures. Our work delimits the applicability
	of large deviation theory and provides methods to probe these limits
	numerically.
\end{abstract}
\maketitle
The fluctuating dynamics of many physical, chemical and biological
systems are commonly modelled by stochastic differential equations
expressed in Langevin or Itô forms \citep{kampenStochasticProcessesPhysics2011,gardinerStochasticMethodsHandbook2010,riskenFokkerPlanckEquationMethods2012,bharucha-reidElementsTheoryMarkov2012}.
In such systems it is often of great interest to identify the typical
pathways that stochastic paths take to transition from an initial
to a final state, as for example in the nucleation of solids, the
conformational changes in biomolecules, or shifts in ecological balance
\citep{faccioliDominantPathwaysProtein2006,demarcoPhaseTransitionModel2001,gardnerConstructionGeneticToggle2000,mangelBarrierTransitionsDriven1994,wolynesNavigatingFoldingRoutes1995,huangMolecularMathematicalBasis2012,paninskiMostLikelyVoltage2006,noltingBallsCupsQuasipotentials2016,leeFindingMultipleReaction2017}.
Typically, such transition paths cluster around multiple pathways
in the space of configurations and the relative probability of one
or the other of these pathways depends on the drift, the diffusivity,
and the duration allowed for the transition to take place \citep{onsagerFluctuationsIrreversibleProcesses1953,bachFunctionalsPathsDiffusion1977,itoProbabilisticConstructionLagrangean1978,ikedaStochasticDifferentialEquations2014}.
As transitions are often rare events, direct simulations are not always
practical and other means, analytical or numerical, are required to
study them. Methods that allow for a full exploration of the space
of transition pathways in stochastic dynamical systems, then, are
of substantial theoretical and practical importance.

The theory of large deviations \citep{ventselSMALLRANDOMPERTURBATIONS1970,stratonovichMarkovMethodsTheory1989,grahamMacroscopicPotentialsBifurcations1989,arnoldStochasticDifferentialEquations1974}
provides an analytical method for obtaining transition pathways -
instantons - in regimes dominated by the drift and for very long durations
of path. Experimental systems, however, are typically not in a regime
where the diffusivity is asymptotically low and durations are asymptotically
long \citep{gladrowExperimentalMeasurementRelative2021}. While the
relevance of including finite-temperature fluctuations around the
instanton \citep{gelfandIntegrationFunctionalSpaces1960} is increasingly
recognized \citep{nickelsen_noise_2022,corazza_normalized_2020,lu_gaussian_2017},
the physical implications of these fluctuations are far from being
understood.

In this Letter, we show that the competition between drift and diffusion
in transition pathways can be studied using semi-classical expansions
of the path measure of the stochastic dynamics. We use a mixture of
Gaussian measures to approximate the path measure around its local
instantons. This allows us to demarcate and transcend the boundaries
of the low diffusivity regime. We demonstrate this explicitly for
a two-dimensional overdamped mechanical system, with both conservative
and non-conservative forces. For this system we uncover a counterintuitive
phenomenon where typical transition paths do not concentrate around
the most probable path, even at low-to-intermediate diffusivities
where the Gaussian approximation is still valid. To validate our results
numerically, we construct a Markov Chain Monte Carlo (MCMC) method
that allows for simultaneous exploration of multiple transition pathways.We
find excellent agreement between the semi-classical expansion and
numerical results for a large range of diffusivities and path durations.
We now detail our results.

\emph{The transition path ensemble. }We consider the stochastic process
generated by the $d$-dimensional overdamped Langevin equation, expressed
in Itô form as

\begin{equation}
d\mathbf{X}=\mu\mathbf{F}dt+\sqrt{2D}d\mathbf{W}.\label{eq:ito equation}
\end{equation}
This represents the stochastic displacement $d\mathbf{X}$ in a time
interval $dt$ of a particle with coordinate $\mathbf{X}$ subject
to a force field $\mathbf{F}$ and Brownian displacements $\sigma d\mathbf{W}$,
where $\mathbf{W}$ is the Wiener process. The particle mobility is
$\mu$, the diffusion constant is $D=\mu/\beta$, and the temperature
is $\theta$ with $\beta^{-1}=k_{B}\theta$, and $k_{B}$ the Boltzmann
constant. We are interested in realisations $\mathbf{X}(t)$ of Eq.~\eqref{eq:ito equation}
that are of duration $T$ and have fixed terminii $\mathbf{X}(0)=\mathbf{x}_{0}$
and $\mathbf{X}(T)=\mathbf{x}_{T}$. These trajectories form a set
of continuous paths that we call the transition path ensemble (TPE).
While in the following we investigate the temperature-dependence of
the TPE for specific model systems, the methods we develop are general.

\begin{figure*}
	\centering
	
	\subfigure[]{{\includegraphics[width=0.31\textwidth]{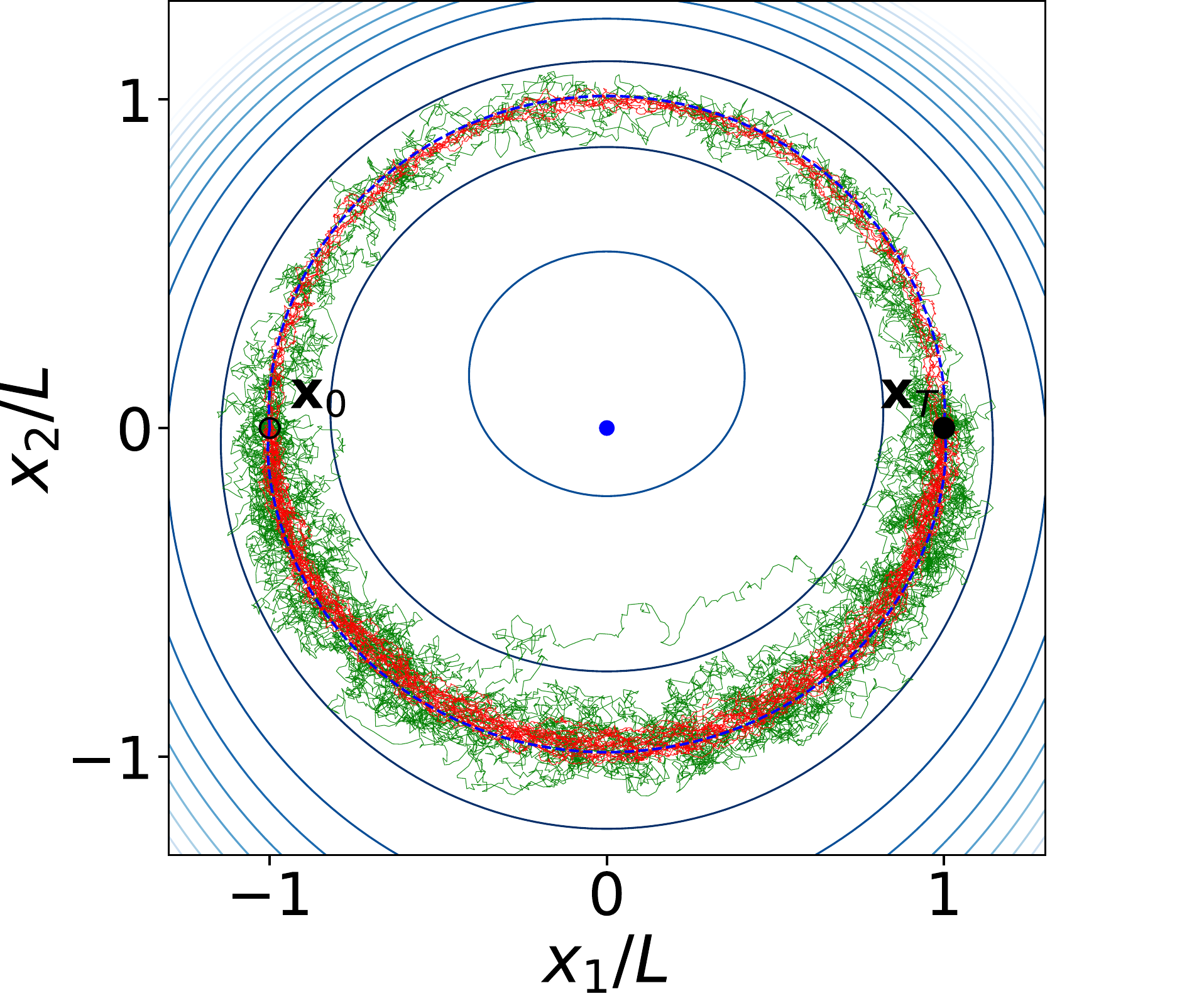}
	}} \hfill{}\subfigure[]{{\includegraphics[width=0.33\textwidth]{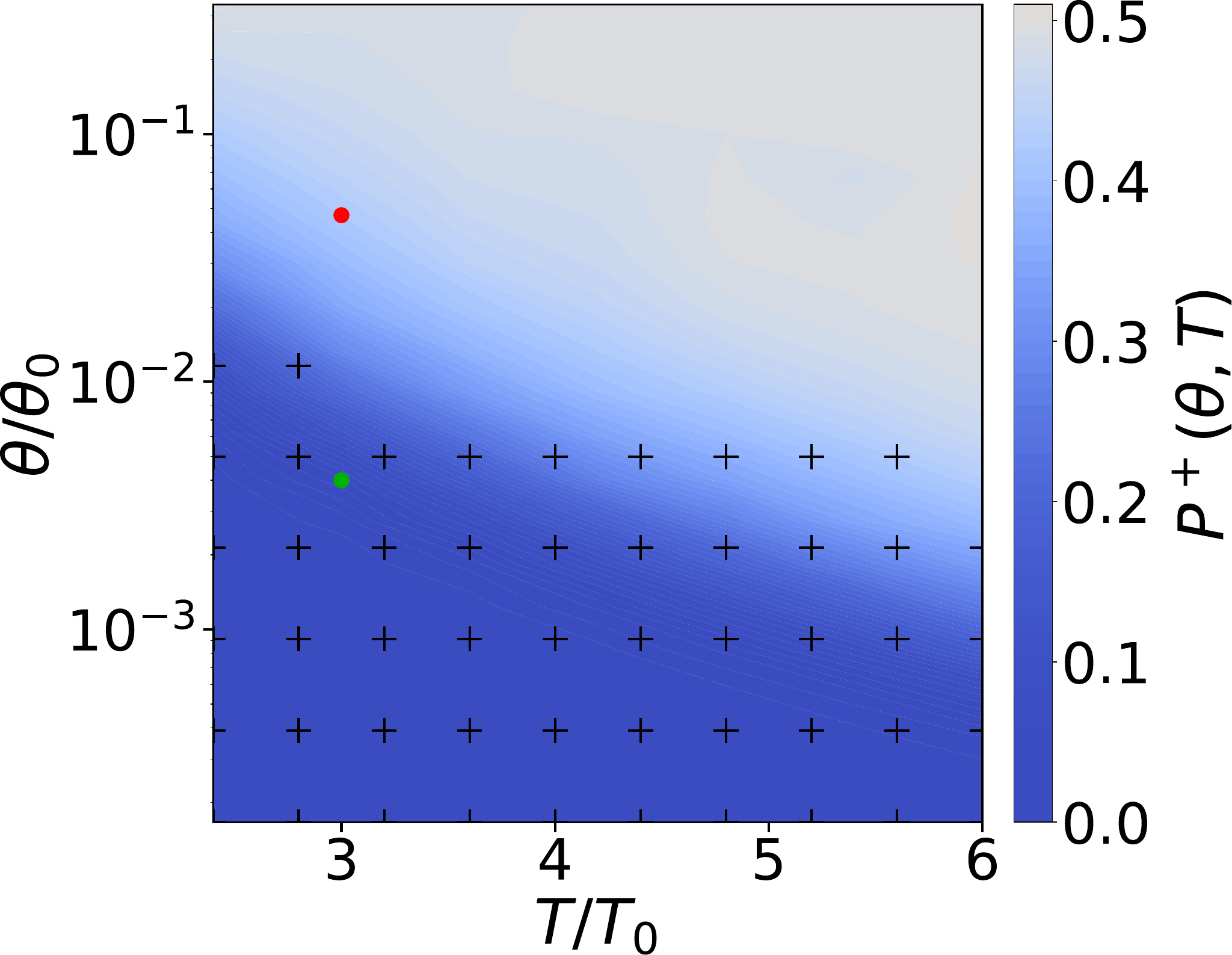}
	}} \hfill{}\subfigure[]{{\includegraphics[width=0.314\textwidth]{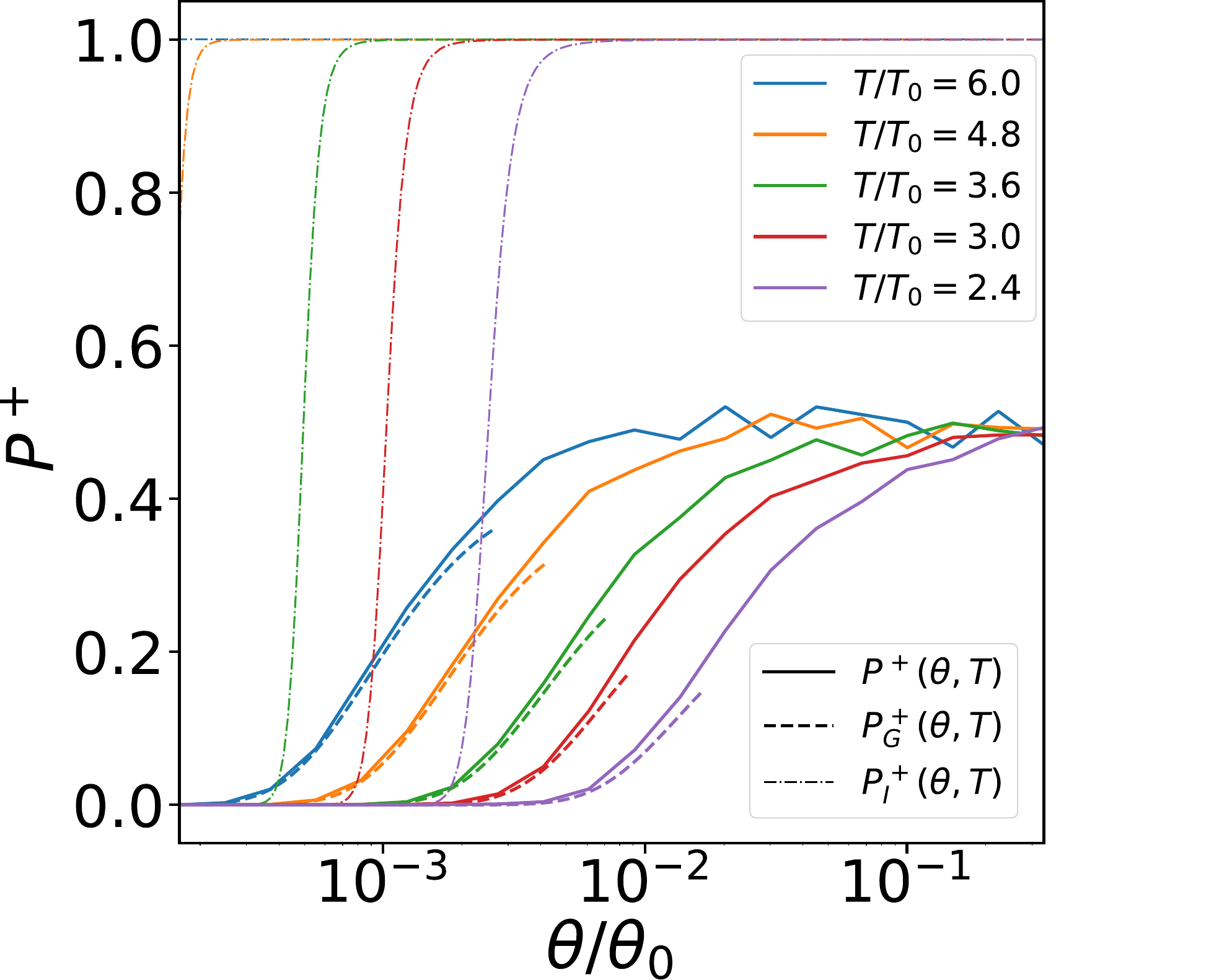}}}%
	\caption{Diffusivity-dependence of the transition path ensemble for the conservative
		model system. Panel (a) shows 50 stochastic trajectories sampled using
		the TMC method (see text) for overdamped dynamics in the potential
		$U(x)$ \citep{note:SI}. The dashed blue (green) lines are upper
		(lower) instantons between initial (circle, $\mathbf{x}_{0}$) and
		final (filled circle, $\mathbf{x}_{T}$) points. Upper and lower channels
		are equally populated at temperature $\theta=0.047\theta_{0}$ (green)
		but the lower channel is preferred at the lower temperature $\theta=0.004\theta_{0}$
		(red). Trajectories of duration $T=3T_{0}$ are sampled with $N=200(T/T_{0})$
		modes. Panel (b) is a pseudocolor plot quantifying the variation of
		the upper and lower channel probabilites with temperature and duration,
		as obtained from TMC. The plus signs show regions where the Gaussian
		mixture approximation $P_{G}^{+}$, defined in Eq. (5), is within
		a $5\%$ margin of error of the simulated value. The red and green
		dots correspond to the same color-coded simulations in panel (a).
		Panel (c) shows a comparision between the TMC, Gaussian mixture and
		instanton approximations to the upper channel probability as function
		of temperature for fixed durations of path. }
	\label{fig:switch}
\end{figure*}

\emph{Model system.} We consider the motion of a particle in $d=2$
dimensions in a potential force field $\mathbf{F}=-\nabla U(\mathbf{x})$;
below we will also add a non-conservative force $\mathbf{F}^{a}$.
The potential $U(\mathbf{x})$ is a deformed Mexican hat, with a maximum
at the origin and a manifold of minima on the circle of radius $L$
around the origin, see Fig.~\ref{fig:switch} (a) for a plot of $U$
and the SI for the explicit parametrisation \citep{note:SI}. We consider
the TPE for paths of duration $T$ which start at $\mathbf{x}_{0}=(-L,0)$
and end at $\mathbf{x}_{T}=(L,0)$. This ensemble features two competing
transition channels, namely along the upper and lower semi-circle,
which we denote by $\Gamma^{+}$ and $\Gamma^{-}$; by design the
potential along $\Gamma^{+}$ is narrower as compared to along $\Gamma^{-}$.

\emph{Gaussian mixture approximation of the TPE.} The TPE is characterized
by its corresponding probability measure $\mathbb{P}$ on the space
of all continuous transition paths. In the path-integral formalism,
this measure is represented by a formal density $\rho[\mathbf{x}(t)]=\exp(-S_{\text{OM}}[\mathbf{x}(t)])$
with respect to a fictitious infinite-dimensional Lebesgue measure
\citep{takahashiProbabilityFunctionalsOnsagermachlup1981}, with the
Onsager-Machlup action \citep{onsagerFluctuationsIrreversibleProcesses1953,bachFunctionalsPathsDiffusion1977,itoProbabilisticConstructionLagrangean1978}

\begin{equation}
S_{\text{OM}}[\mathbf{x}(t)]=\int_{0}^{T}\left(\frac{\beta}{4\mu}|\dot{{\bf x}}-\mathbf{F}|^{2}+\frac{\mu}{2}\nabla\cdot\mathbf{F}\right)dt.\label{eq:onsager-machlup action}
\end{equation}

The variational minima of the action Eq.~\eqref{eq:onsager-machlup action}
have physical relevance. Namely, because the first term in Eq.~\eqref{eq:onsager-machlup action}
is inversely proportional to the temperature, at sufficiently low
temperature the formal probability density $\rho$ is dominated by
paths that aggregate around the variational minima \citep{ventselSMALLRANDOMPERTURBATIONS1970,stratonovichMarkovMethodsTheory1989,grahamMacroscopicPotentialsBifurcations1989}.
These variational minima are called the local instantons of the Onsager-Machlup
action, and we denote them by $\mathbf{x}^{[\alpha]}(t;\ \theta,T)$,
with $\alpha=1,\dots,K$, $\mathbf{x}^{[\alpha]}(0;\ \theta,T)=\mathbf{x}_{0}$
and $\mathbf{x}^{[\alpha]}(T;\ \theta,T)=\mathbf{x}_{T}$. The arguments
of the local instantons indicate that they depend on the temperature
and the duration of the path (we suppress these arguments below).
For the potential $U$ from Fig.~\ref{fig:switch} (a) we find $K=2$
local instantons $\mathbf{x}^{[1]}\equiv\mathbf{x}^{+}$, $\mathbf{x}^{[2]}\equiv\mathbf{x}^{-}$,
going along the upper and lower semi-circles, respectively.

By performing a quadratic functional Taylor expansion of Eq.~\eqref{eq:onsager-machlup action}
around the $\alpha$-th local instanton \citep{engelDensityFunctionalTheory2011,gelfandCalculusVariations2012,corazzaNormalizedGaussianPath2020a},

\begin{equation}
S_{\text{OM}}[\mathbf{x}(t)]\approx S_{\text{OM}}[\mathbf{x}^{[\alpha]}(t)]+\frac{1}{2}\langle\delta\mathbf{x},\mathcal{H}^{[\alpha]}\delta\mathbf{x\rangle},\label{eq:quadratic expansion}
\end{equation}
where $\mathcal{H}^{[\alpha]}$ is a self-adjoint linear differential
operator, $\delta\mathbf{\mathbf{x}}=\mathbf{x}-\mathbf{x}^{[\alpha]}$
and $\langle\mathbf{f},\mathbf{g}\rangle=\sum_{i}\int_{0}^{T}f_{i}(t)g_{i}(t)dt$,
we can formally define a Gaussian measure $\mathbb{P}^{[\alpha]}$
with mean $\mathbf{x}^{[\alpha]}$, precision $\mathcal{H}^{[\alpha]}$,
\textit{\emph{and regularised normalisation constant $\mathcal{Z}^{[\alpha]}$,
}}in the space of paths. The resulting $K$ local approximators of
the measure can be combined into a Gaussian mixture approximation
\textit{\emph{\citep{gelmanBayesianDataAnalysis}}} of the whole TPE
\begin{equation}
\mathbb{P}\approx\bar{\mathbb{P}}\equiv\sum_{\alpha=1}^{K}w_{\alpha}\mathbb{P}^{[\alpha]},\label{eq:gaussian mixture approximation of TPE}
\end{equation}
where the weights $w_{\alpha}=e^{-S_{\text{OM}}[\mathbf{x}^{[\alpha]}]}\mathcal{Z}^{[\alpha]}/\sum_{\gamma=1}^{K}e^{-S_{\text{OM}}[\mathbf{x}^{[\gamma]}]}\mathcal{Z}^{[\gamma]}$
satisfy $\sum_{\alpha=1}^{K}w_{\alpha}=1$, see the SI \citep{note:SI}
for more details. Equation \eqref{eq:gaussian mixture approximation of TPE}
is the infinite-dimensional analogue to approximating a finite-dimensional
multimodal probability density by a sum of Gaussians, with one term
for each local maximum of the probability density.

\textit{Transition channel probabilities.}\textit{\emph{ For temperature
		$\theta$ and path duration $T$ we define $P^{[\alpha]}(\theta,T)$
		as the probability of observing a transition path travelling via the
		$\alpha$-th channel, i.e.}}$~$\textit{\emph{close to the $\alpha$-th
		instanton. Using Eq.}}~\textit{\emph{\eqref{eq:gaussian mixture approximation of TPE}
		we approximate $P^{[\alpha]}(\theta,T)$ as}}
\begin{equation}
P^{[\alpha]}(\theta,T)\approx P_{G}^{[\alpha]}(\theta,T)\equiv\frac{e^{-S_{\text{OM}}[\mathbf{x}^{[\alpha]}]}\mathcal{Z}^{[\alpha]}}{\sum_{\gamma=1}^{K}e^{-S_{\text{OM}}[\mathbf{x}^{[\gamma]}]}\mathcal{Z}^{[\gamma]}}.\label{eq:r function}
\end{equation}
\textit{\emph{ According to Eq.}}~\textit{\emph{\eqref{eq:r function}
		the relative channel probabilities are determined by an interplay
		between the instanton probabilities, as quantified by $e^{-S_{\text{OM}}[\mathbf{x}^{[\alpha]}]}$,
		and the sizes of the Gaussian fluctuations around the local instantons,
		$\mathcal{Z}^{[\alpha]}$. It is instructive to compare this ratio
		with another estimator $P_{I}^{[\alpha]}(\theta,T)=e^{-S_{\text{OM}}[\mathbf{x}^{[\alpha]}]}/\sum_{\gamma}e^{-S_{\text{OM}}[\mathbf{x}^{[\gamma]}]}$,
		in which only the instanton probabilities are retained.}}

To use Eq.~\eqref{eq:r function} in practice, we retrieve the instantons
$\mathbf{x}^{[\alpha]}$ using a Ritz variational method presented
in \citep{kikuchiRitzMethodTransition2020,gladrowExperimentalMeasurementRelative2021}.\textit{\emph{
}}We subsequently evaluate the regularised normalisation constants
\textit{\emph{$\mathcal{Z}^{[\alpha]}$}} using the Gelfand-Yaglom
theorem \citep{dunneFunctionalDeterminantsQuantum2008,gelfandIntegrationFunctionalSpaces1960,levitTheoremInfiniteProducts1977,corazzaNormalizedGaussianPath2020a},
as well as a generalisation thereof to non-gradient dynamics which
we provide in the SI \citep{note:SI}.

\textit{Numerical experiments.}\textit{\emph{ To infer the range of
		validity of our semi-analytical approximation it is necessary to compare
		Eq.~\eqref{eq:r function} with numerical simulations. In parameter
		regimes where transitions are very rare, it is not feasible to sample
		the TPE using direct simulations. We therefore numerically probe the
		TPE using a MCMC algorithm built on the }}\textit{preconditioned Crank-Nicholson
	algorithm }\textit{\emph{(pCN) }}\citep{cotterMCMCMethodsFunctions2013,beskosMCMCMETHODSDIFFUSION2008,hairerSpectralGapsMetropolis2014},
as detailed in the SI \citep{note:SI}. In essence, we approximate
the function space of all transition paths by a finite sum of basis
functions \citep{kosambiStatisticsFunctionSpace1943,karhunenUberLineareMethoden1947,loeveProbabilityTheory1977},
and perform a random walk on the resulting finite-dimensional space
of basis coefficients; the random walk is designed such that the resulting
transition path samples are distributed according to the TPE we seek
to probe.

A general shortcoming of MCMC methods and also other transition path
sampling techniques \citep{bolhuisTransitionPathSampling2002,bolhuisTransitionPathSampling,dellagoTransitionPathSampling1998,dellagoEfficientTransitionPath1998a,fujisakiOnsagerMachlupActionbased2010}
is that when the distribution to be sampled is multimodal with regions
of low probability in between the modes, it may take prohibitively
long to obtain converged results. For overdamped Langevin dynamics
Eq.~\eqref{eq:ito equation}, this corresponds to medium-to-low temperature
regimes in systems with competing transition pathways, where the TPE
concentrates around the local instantons. One way to overcome this
issue is to use replica exchange \citep{fujisakiOnsagerMachlupActionbased2010},
which requires running several instances of the MCMC algorithm at
varying temperatures. Here we introduce a modification of the pCN-MCMC
that operates only at one temperature, which we call the \emph{Teleporter
	MCMC }(TMC), which utilises the Gaussian mixture approximation of
the TPE. At each step of the TMC there is a small probability to jump
between the transition channels, which accelerates mixing between
them. We provide a detailed description of the algorithm in the SI
\citep{note:SI}.

\textit{Results.} We now consider the transition behavior of the 2D
system depicted in Fig.~\ref{fig:switch} (a). For a range of temperatures
$\theta$ and total transition times $T$ we first generate ensembles
of $10^{8}$ sample transition paths per tuple $(\theta,T)$ using
the TMC. Let $\tau_{D}(\theta)=L^{2}/(\mu k_{B}\theta)$, which is
the diffusive time-scale at temperature $\theta$. We also introduce
fixed reference temperature and time-scales $\theta_{0}=U_{0}/k_{B}$
and $T_{0}=\tau_{D}(\theta_{0})$, where $U_{0}$ is the energetic
well-depth of the potential. Our parameter range is such that $T\ll\tau_{D}$,
for each temperature $\theta$ in the range considered. Each sampled
ensemble thus describes a rare transition event. For a total transition
time $T/T_{0}=3$, and for each of the two temperatures $\theta/\theta_{0}=0.047$
and $\theta/\theta_{0}=0.004$, we show 50 randomly chosen TMC sample
paths in Fig.~\ref{fig:switch} (a).  We observe that while for the
higher temperature the paths are evenly distributed between the two
channels, for the lower temperature the lower channel is preferred.
In Fig.~\ref{fig:switch} (b) we show TMC results for $P^{+}(\theta,T)\equiv P^{[1]}(\theta,T)$,
the probability of the upper channel, as a function of both $\theta$
and $T$. Consistent with the $\theta/\theta_{0}=0.047$ data from
Fig.~\ref{fig:switch} (a), we observe that for large enough temperature
$P^{+}(\theta,T)\approx1/2$ (white region), so that upper and lower
channel are equally probable. That at large temperature the asymmetry
in $U$ becomes irrelevant for the TPE is expected, as in this limit
the random force in Eq.~\eqref{eq:ito equation} dominates over the
deterministic force. As $\theta$ is decreased, the channel around
$\Gamma^{-}$ becomes dominant, so that $P^{+}(\theta,T)\rightarrow0$
(blue region in Fig.~\ref{fig:switch} (b), c.f.~$\theta/\theta_{0}=0.004$
data in subplot (a)). The exact temperature at which the crossover
from the diffusivity-dominated regime to the drift-dominated regime
occurs decreases with increasing $T$; this is clearly seen in Fig.~\ref{fig:switch}
(c) where vertical sections of subplot (b) are shown for several values
of $T$.

We now compare our numerical TMC results for $P^{+}(\theta,T)$ with
the Gaussian mixture approximation $P_{G}^{+}(\theta,T)\equiv P_{G}^{[1]}(\theta,T)$,
defined in Eq.~\eqref{eq:r function}. Figure \ref{fig:switch} (b)
shows that this approximation is valid in the low-temperature regime
(plus signs). This is consistent with the assumptions underlying the
Gaussian approximation, as we expect the probability distribution
in path space to be dominated by the neighborhoods of the local instantons
only for sufficiently low temperature. As Fig.~\ref{fig:switch}
(c) shows, $P_{G}^{+}(\theta,T)$ quantitatively captures the beginning
of the crossover from drift-dominated to diffusivity-dominated transition
behaviour for all values of $T$ considered.

For capturing this $\theta$-dependent crossover, the prefactors $\mathcal{Z}^{+}=\mathcal{Z}^{[1]}$,
$\mathcal{Z}^{-}=\mathcal{Z}^{[2]}$ in Eq.~\eqref{eq:r function}
are essential. This becomes apparent by considering $P_{I}^{+}(\theta,T)$,
which only depends on the relative probabilities of the two local
instantons. In Fig.~\ref{fig:switch} (c) we see that for high enough
temperatures $P_{I}^{+}(\theta,T)\approx1$, meaning $S_{\text{OM}}[\mathbf{x}^{+}(t)]<S_{\text{OM}}[\mathbf{x}^{-}(t)]$
\citep{adibStochasticActionsDiffusive2008}. This limit is understood
by comparing the two terms in the action Eq.~\eqref{eq:onsager-machlup action}.
While the first term scales as $1/\theta$, the second term is independent
of $\theta$; for fixed $T$ and large enough $\theta$ the second
term thus dominates the action. This second term is smaller for the
channel around $\Gamma^{+}$ than for the channel around $\Gamma^{-}$,
because the former channel is narrower leading to to a smaller value
of $\nabla\cdot\mathbf{F}$. As $\theta$ is decreased for fixed $T$
the first term in Eq.~\eqref{eq:onsager-machlup action} becomes
dominant. Figure \ref{fig:switch} (c) shows that this leads to a
crossover to $P_{I}^{+}(\theta,T)\approx0$, meaning $\mathbf{x}^{-}$
becomes more probable than $\mathbf{x}^{+}$. While this low-temperature
limit is consistent with the numerical results, the temperature at
which we observe the crossover in $P_{I}^{+}(\theta,T)$ is smaller
as compared to $P^{+}(\theta,T)$. For example, we see in Fig.~\ref{fig:switch}
(c) that for $T/T_{0}=2.4$ the crossover of $P_{I}^{+}(\theta,T)$
is at $\theta/\theta_{0}<10^{-2}$, whereas the crossover for $P^{+}(\theta,T)$
occurs at $\theta/\theta_{0}>10^{-2}$. In particular this implies
that for $\theta/\theta_{0}=10^{-2}$ the most probable path goes
along $\Gamma^{+}$, while most transition paths go along $\Gamma^{-}$.
This highlights that even at intermediate-to-low temperatures, where
the Gaussian mixture approximation Eq.~\eqref{eq:r function} is
already valid, the probabilities of the local instantons alone are
insufficient to obtain the actual transition behaviour. Instead it
is the prefactors $\mathcal{Z^{\pm}}$ in Eq.~\eqref{eq:r function}
that dominate the crossover behaviour in Fig.~\ref{fig:switch} (b);
these Gaussian normalisation constants are, in a sense, an entropic
contribution, as they measure the effective volume in path space of
the support around the respective local instanton. Even though for
$T/T_{0}=2.4$, $\theta/\theta_{0}=10^{-2}$ the instanton $\mathbf{x}^{+}$
is more probable than $\mathbf{x}^{-}$, this is more than offset
by the larger number of paths that behave similar to $\mathbf{x}^{-}$.
As we discuss in the SI \citep{note:SI}, the prefactors $\mathcal{Z^{\pm}}$
remain relevant even in the Freidlin-Wentzell-Graham limit \citep{ventselSMALLRANDOMPERTURBATIONS1970,grahamMacroscopicPotentialsBifurcations1989}
of vanishing temperature and infinite path duration.

\begin{figure}[t]
	\includegraphics[width=0.36\textwidth]{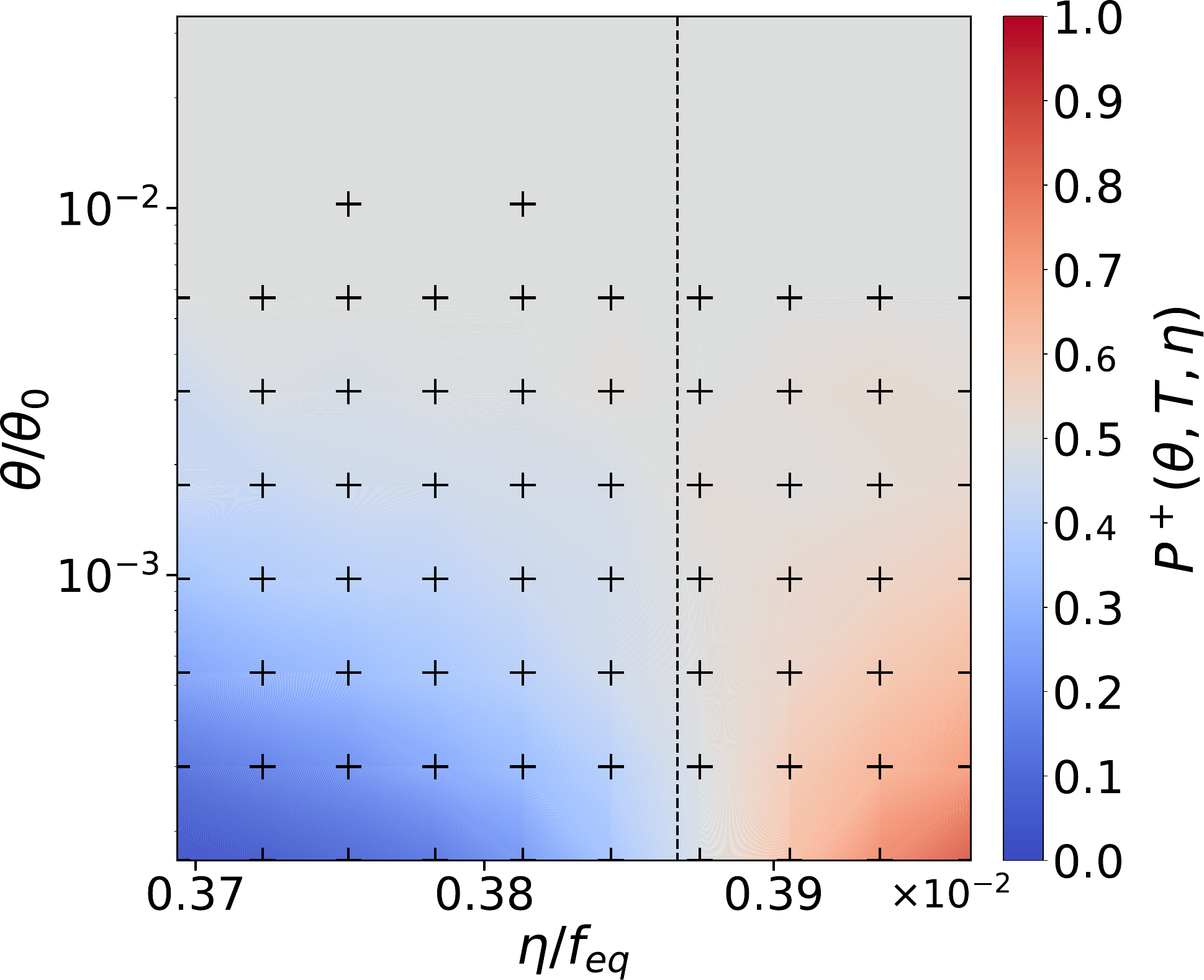}
	\centering \caption{Diffusivity-dependence of the transition path ensemble for the non-conservative
		model system. Pseudocolour plot of the probability of the upper channel,
		$P^{+}(\theta,T,\eta)$, for the non-gradient system with $T/T_{0}=3$,
		as a function of the temperature $\theta$ and the circular force-strength
		$\eta$. The black plus signs show regions where the variational approximation
		$P_{G}^{+}(\theta,T,\eta)$, defined in Eq. \eqref{eq:r function},
		is within a 5\% margin of error of the simulated value. The dashed
		line shows the crossover force strength $\eta_{c}/f_{\text{eq}}\approx0.00387$,
		where $f_{\text{eq}}$ is the characteristic strength of the gradient
		force \citep{note:SI}.}
	\label{fig:switch noneq}
\end{figure}

For non-gradient forms of the drift, the prefactors $\mathcal{Z^{\pm}}$
can also drive the crossover behaviour of the system, as we show now
by adding a force of strength $\eta$ that acts perpendicular to the
radius vector in the clockwise direction. For positive force strength
$\eta$, this non-conservative force biases towards the upper channel
$\Gamma^{+}$. In Fig.~\ref{fig:switch noneq} we show numerical
results for $P^{+}(\theta,T,\eta)$ as a function of $\eta$ and $\theta$
for $T/T_{0}=3$. For small $\eta/f_{\text{eq}}\rightarrow0$, with
$f_{\text{eq}}$ the characteristic strength of the gradient force
\citep{note:SI}, we recover the results from Fig.~\ref{fig:switch}
(b), (c). Thus at small but finite temperature the dominant transition
channel is the one where particles travel againt the weak non-conservative
force. As $\eta$ is increased to $\eta_{c}/f_{\text{eq}}\approx0.00387$,
we observe a crossover from $\Gamma^{-}$-channel dominated transitions
to $\Gamma^{+}$-channel dominated transitions in the low-temperature
regime. This switch is also captured by the Gaussian approximation
(plus signs in Fig.~\ref{fig:switch noneq}). On the other hand,
throughout the parameter regime considered in Fig.~\ref{fig:switch noneq},
we find that $P_{I}^{+}(\theta,T,\eta)\approx1$, meaning that the
local instanton $\mathbf{x}^{+}$ is always more probable than $\mathbf{x}^{-}$
for finite $\eta$. This again highlights the relevance of considering
Gaussian fluctuations around the instantons for determining the dominant
transition pathway.

\textit{Conclusion. }For a system with two competing transition pathways,
we have studied how the dominant transition pathway depends on both
the temperature and the total duration. To quantify the relative importance
of the competing pathways, we have constructed semi-analytical approximators
which are valid in the low-to-intermediate temperature regime. We
have validated our approximators via comparison with a continuous-time
MCMC method that is dimensionally robust and efficiently samples systems
with multiple reactive pathways.

Our results show that even in the low-to-intermediate temperature
regime the global instanton, or most probable path, itself is not
sufficient to determine the dominant transition pathway. Rather, it
is vital that fluctuations around this path be incorporated. This
has a simple one-dimensional equivalent: For a probability density
$\rho(x)\sim\exp(-V(x))$ for some potential $V(x)$ with relative
minima $x_{\alpha}$, the probabilistically most relevant minimum
is not the global one, but that with the largest well probability,
i.e.$~$the $x_{\alpha}$ that maximizes $P(\,\mathrm{well~around~}x_{\alpha}\,)\sim e^{-V(x_{\alpha})}\sqrt{{2\pi}/{V''(x_{\alpha})}}$,
where we use a quadratic Taylor approximation of $V$ around $x_{\alpha}$.
The most probable well is thus determined by an interplay of $e^{-V(x_{\alpha})}$
(which corresponds to the instanton probability $e^{-S_{\text{OM}}[\mathbf{x}^{[\alpha]}]}$)
and $\sqrt{{2\pi}/{V''(x_{\alpha})}}$ (which corresponds to the regularised
normalisation constant $\mathcal{Z}^{[\alpha]}$).

In the present paper we consider a paradigmatic example system with
two competing transition pathways. The method of instantons is an
established technique in theoretical chemistry and statistical physics
\citep{eMinimumActionMethod2004,e_string_2002,grafke_instanton_2015,grafke_numerical_2019,schorleppGelFandYaglomType2021,dematteis_experimental_2019,ferreApproximateOptimalControls2021,kikuchiRitzMethodTransition2020,heymannGeometricMinimumAction2008},
and the method of Gaussian mixtures presented here scales as $O(d^{2})$
with the number of degrees of freedom $d$ \citep{note:SI}. It is
therefore feasible to apply the methods we developed here to more
realistic many-particle systems to study e.g.$~$nucleation pathways
\citep{e_string_2002,lutskoHowCrystalsForm2019,rein_ten_wolde_numerical_1996}
or conformational rearrangements in macromolecules \citep{ren_transition_2005,fujisakiOnsagerMachlupActionbased2010,fujisakiMultiscaleEnhancedPath2013,gartner_modeling_2019}.

Our quantification here of the finite-temperature breakdown of instanton
theories is important for relating such theories to experiments, which
are always at finite temperatures. Our insights into path-space probability
distributions for diffusive stochastic dynamics, together with our
MCMC method, will therefore be valuable for going beyond the regime
of asymptotically low diffusivity in both large deviations theory
\citep{ventselSMALLRANDOMPERTURBATIONS1970,vanden-eijndenGeometricMinimumAction2008,eMinimumActionMethod2004,grafke_numerical_2019}
and the study of rare events \citep{grafkeSharpAsymptoticEstimates2021}.

\begin{acknowledgments} \textit{Acknowledgements.} We thank Professor M. E. Cates for many
	helpful discussions. Work funded in part by the European Research
	Council under the Horizon 2020 Programme, ERC grant agreement number
	740269, and by the Royal Society through grant RP1700. \end{acknowledgments}

\onecolumngrid
\appendix

\section{System specification}

Here we define the two systems in consideration in the main text.
We construct a Sombrero-type potential such that the perpendicular
curvature of the potential along the minimal upper semi-circle $\Gamma^{+}$
is larger than that along the lower semi-circle $\Gamma^{-}$. We
start with a radial quartic potential of the form 
\begin{align}
U(x_{1},x_{2}) & =U_{r}(r(x_{1},x_{2}))\\
U_{r}(r) & =U_{0}\left(\frac{r}{L}-1\right)\left(1+a\frac{r}{L}+b\left(\frac{r}{L}\right)^{2}+c\left(\frac{r}{L}\right)^{3}\right)\nonumber 
\end{align}
where $L$ is the length-scale of the system, $U_{0}$ will be the
value of the potential at the local maximum $r=0$, and $a,b,c\in\mathbb{R}$
will be specified below. We will henceforth supress the argument of
the radial coordinate function $r(x_{1},x_{2})=\sqrt{x_{1}^{2}+x_{2}^{2}}$.
Let $\Gamma=\{(x,y)\ |\ r=1\}$ be the circle centered around the
origin, which satisfies $U_{r}(1)=0$. We also define $\Gamma^{+}$and
$\Gamma^{-}$ as the upper and lower semi-circle respectively. We
impose the following conditions on the potential to fix $a,b,c$:
\begin{enumerate}
	\item $U_r'(0) = 0$. The origin is an extremum.
	\item $U_r'(1) = 0$. $\Gamma$ is an extremum of the potential. \item $U_r''(1) = k$. The curvature along $\Gamma$ is $k$.
\end{enumerate}We get
\begin{equation}
U_{r}(r)=\frac{1}{2}\left(\frac{r}{L}-1\right)^{2}\left[L^{2}k\left(\frac{r}{L}\right)^{2}-2U_{0}\left(\frac{r}{L}-1\right)\left(3\frac{r}{L}+1\right)\right]
\end{equation}
In order to ensure that the potential has a Sombrero-like form, we
must further have that the potential is confining, which is equivalent
to $\underset{r\to\infty}{\lim}U_{r}(r)=\infty$, which implies that
$6U_{0}\leq L^{2}k$. We now introduce an angular dependence in the
curvature. We set
\begin{equation}
L^{2}k(\phi)=6U_{0}(1+2h(\phi))\label{eq:curvature}
\end{equation}
where $\phi=\phi(x_{1},x_{2})$ is the angle of $(x_{1},x_{2})$ in
polar coordinates so that $x_{1}=\cos(\phi)$, $x_{2}=\sin(\phi)$,
and where
\begin{equation}
h(\phi)=\frac{1}{4}\left(\xi_{2}+\xi_{1}+(\xi_{2}-\xi_{1})\sin\phi\right)
\end{equation}
where $\xi_{2}>\xi_{1}$, and where $h(\phi)\in[\xi_{1},\xi_{2}]$
satisfies $h(-\pi/2)=\xi_{1}$ and $h(\pi/2)=\xi_{2}$. Eq.~\eqref{eq:curvature}
is constructed so that the perpendicular curvature of $\Gamma^{+}$
is larger than that of $\Gamma^{-}$. The drift of the system is now
given by $\mathbf{F}=-\nabla U$.

For the non-gradient system, we introduce an additional non-conservative
force $\mathbf{F}^{a}=-\eta\hat{\boldsymbol{\phi}}$ for which the
work done in a displacement $d\mathbf{x}=dr\hat{\mathbf{r}}+rd\phi\hat{\boldsymbol{\phi}}$
is $dW=\mathbf{F}^{a}\cdot d\mathbf{x}=\eta rd\phi$. This force energetically
biases the upper transition channel $\Gamma^{+}$. The total force
is thus $\mathbf{F}=-\nabla U+\mathbf{F}^{a}$.

In the numerical experiments presented in the main text, we used the
Itô Langevin equation 

\begin{equation}
d\mathbf{X}=\mu\mathbf{F}dt+\sqrt{2\mu k_{B}\theta}d\mathbf{W}.\label{eq:langevin equation}
\end{equation}
We now put Eq. \ref{eq:langevin equation} in non-dimensionalised
form by introducing the time-scale $T_{0}=\frac{L^{2}}{k_{B}\theta_{0}\mu}$
and temperature-scale $\theta_{0}$, and setting $t=T_{0}\tilde{t}$,
$\theta=\theta_{0}\tilde{\theta}$, $\mathbf{X}=L\tilde{\mathbf{X}}$,
$\mathbf{F}=\frac{U_{0}}{L}\tilde{\mathbf{F}}$ and $\mathbf{W}=\sqrt{T_{0}}\tilde{\mathbf{W}}.$
$T_{0}$ is the typical diffusion time-scale at temperature $\theta_{0}$.
We get
\[
d\mathbf{\tilde{X}}=\tilde{U}_{0}\tilde{\mathbf{F}}d\tilde{t}+\sqrt{2\tilde{\theta}}d\tilde{\mathbf{W}}.
\]
where $\tilde{U}_{0}=\frac{U_{0}}{k_{B}\theta_{0}}$ is the ratio
of the well-depth $U_{0}$ and the thermal energy at temperature $\theta_{0}$.
For the numerical experiments in the main text  we use $\tilde{U}_{0}=1$,
which means that $\tilde{\theta}=1$ corresponds to a temperature
such that $k_{B}\theta=U_{0}$. We also set $\xi_{1}=0$ and $\xi_{2}=2$.
To compare the gradient force with $\mathbf{F}^{a}$ we also introduce
$f_{\text{eq}}=U_{0}/L$, which is the characteristic force strength
of the gradient force.

\section{Gaussian mixture approximation of the transition path ensemble}

Here we derive an approximation of the transition path ensemble, using
a Gaussian mixture approximation of the path-space probability measure.
The subsequent sections give detailed descriptions of the mathematical
techniques necessary for the approximation, but we will first give
some intuition by drawing an analogy to a one-dimensional probability
density.

For a one-dimensional probability density $\rho(x)=\mathcal{N}^{-1}\exp(-V(x))$,
where $\mathcal{N}$ is a normalization constant and where the potential
$V(x)$ has well-separated relative minima $x_{\alpha},\ \alpha=1,\dots,K$,
we can approximate $\rho(x)$ around $x_{\alpha}$ using a Gaussian
approximation
\begin{equation}
\rho(x)\approx\frac{1}{\mathcal{{N}}}e^{-V(x_{\alpha})-V''(x_{\alpha})(x-x_{\alpha})^{2}/2}=:\frac{{{\mathcal{N}}_{\alpha}}}{\mathcal{{N}}}e^{-V(x_{\alpha})}\rho^{[\alpha]}(x)\label{eq:local approximation of rho}
\end{equation}
with a normalised Gaussian distribution $\rho^{[\alpha]}(x):=\mathcal{N}_{\alpha}^{-1}e^{-V''(x_{\alpha})(x-x_{\alpha})^{2}/2}$
and where $\mathcal{N}_{\alpha}=\sqrt{2\pi/V''(x_{\alpha})}$. Equation
\ref{eq:local approximation of rho} is a local approximation of $\rho(x)$
around $x_{\alpha}$. If $\rho(x)$ is highly peaked around its maxima
(for example if $V(x)$ describes a Boltzmann distribution $V(x)=U(x)/(k_{B}\theta)$
at a low temperature $\theta$), a global approximation of $\rho(x)$
is the Gaussian mixture
\begin{equation}
\rho(x)\approx\sum_{\alpha=1}^{K}\frac{{\mathcal{{N}}}_{\alpha}}{\mathcal{{N}}}e^{-V(x_{\alpha})}\rho^{[\alpha]}(x)=:\sum_{\alpha=1}^{K}w_{\alpha}\rho^{[\alpha]}(x)\label{eq:one_dimensional_sum_of_gaussians}
\end{equation}
where $w_{\alpha}=e^{-V(x_{\alpha})}\mathcal{N}_{\alpha}/\mathcal{{N}}$
are constants that weight the local Gaussian distributions, and where
$\mathcal{N}\approx\sum_{\gamma=1}^{K}e^{-V(x_{\gamma})}\mathcal{N}_{\gamma}$. 

Equation \eqref{eq:one_dimensional_sum_of_gaussians} can be used
to approximately evaluate any expectation value. In particular, the
probability of being in well $\alpha$ (i.e.~around $x_{\alpha}$)
is given by

\begin{align}
P(x\in\text{{well\,}}\alpha) & =\mathbb{E}[\chi_{\alpha}]=\int_{-\infty}^{\infty}dx\,\chi_{\alpha}(x)\rho(x)\approx\sum_{\beta=1}^{K}w_{\beta}\int_{-\infty}^{\infty}dx\,\chi_{\alpha}(x)\rho^{[\beta]}(x)\\
& \approx w_{\alpha}\int_{-\infty}^{\infty}dx\,\rho^{[\alpha]}(x)=w_{\alpha}=\frac{{e^{-V(x_{\alpha})}\mathcal{N}_{\alpha}}}{\sum_{\gamma=1}^{L}e^{-V(x_{\gamma})}\mathcal{N}_{\gamma}},\label{eq:one_dimensional_well_probability}
\end{align}
where the indicator function $\chi_{\alpha}(x)$ is 1 if $x$ is in
well $\alpha$ and zero otherwise, and where we assume that the potential
wells of $V(x)$ are well-separated so that $\chi_{\alpha}(x)\rho^{[\beta]}(x)$
is negligibly small whenever $\alpha\neq\beta$.

In the following we apply the same steps as above to the case of the
transition path ensemble (TPE). As we are considering Gaussian approximations
of probability distributions over infinite-dimensional functional
spaces, the mathematical sophistication required is higher, but the
intuition remains identical to the above one-dimensional example.
In Sec.~II.A we derive the local Gaussian approximation for path-probability
measures around an instanton (which is based on a second-order functional
Taylor approximation), which we in Sec.~II.B combine to a Gaussian
mixture approximation of the TPE. In Sec.~II.D we derive the method
we use to calculate the normalisation constants for functional Gaussians
(i.e. the infinite-dimensional equivalent of the $\mathcal{N}_{\alpha}$
from the one-dimensional example above). We put Sec.~II.D last as
it is a technical result, and is not required to understand the rest
of the subsections. In Sec.~II.C we use the Gaussian mixture to derive
an approximate expression for transition pathway probabilities, which
proceeds analogous to the calculation Eq.~\eqref{eq:one_dimensional_well_probability}.

\subsection{The quadratic expansion of the Onsager-Machlup action}

Here we describe how to formally construct the Gaussian expansion
around a given reference path. The variational expansion of the Onsager-Machlup
action

\begin{align}
S_{\text{OM}}[\mathbf{x}(t)] & =\int_{0}^{T}L(\mathbf{x}(t),\dot{\mathbf{x}}(t))dt\label{eq:onsager-machlup action}\\
L(\mathbf{x},\dot{\mathbf{x}}) & =\frac{\beta}{4\mu}|\dot{{\bf x}}-\mathbf{F}|^{2}+\frac{\mu}{2}\nabla\cdot\mathbf{F}\nonumber 
\end{align}
where $\beta=1/k_{B}\theta$ is the inverse temperature, is given
by
\begin{align}
S_{\text{OM}}[\mathbf{\mathbf{\bar{\mathbf{x}}}}+\delta\mathbf{x}] & =S_{\text{OM}}[\mathbf{\mathbf{\bar{\mathbf{x}}}}]+\mathrm{J}[\delta\mathbf{x}]+\frac{1}{2}\mathrm{H}[\delta\mathbf{x}]+O(\delta\mathbf{x}^{3})\label{eq:variation of OM}
\end{align}
to second order around a reference path $\mathbf{\bar{\mathbf{x}}}(t)$,
where
\begin{align}
\mathrm{J}[\delta\mathbf{x}] & =\int_{0}^{T}\left\{ \frac{\partial L}{\partial\mathbf{x}}(\mathbf{\mathbf{\bar{\mathbf{x}}}},\dot{\mathbf{\mathbf{\bar{\mathbf{x}}}}})\cdot\delta\mathbf{x}+\frac{\partial L}{\partial\dot{\mathbf{x}}}(\mathbf{\mathbf{\bar{\mathbf{x}}}},\dot{\mathbf{\mathbf{\bar{\mathbf{x}}}}})\cdot\delta\dot{\mathbf{x}}\right\} dt\\
\mathrm{H}[\delta\mathbf{x}] & =\int_{0}^{T}\left\{ \delta\mathbf{x}\cdot\frac{\partial^{2}L}{\partial\mathbf{x}\partial\mathbf{x}}(\mathbf{\mathbf{\bar{\mathbf{x}}}},\dot{\mathbf{\mathbf{\bar{\mathbf{x}}}}})\cdot\delta\mathbf{x}+2\ \delta\mathbf{x}\cdot\frac{\partial^{2}L}{\partial\mathbf{x}\partial\dot{\mathbf{x}}}(\mathbf{\mathbf{\bar{\mathbf{x}}}},\dot{\mathbf{\mathbf{\bar{\mathbf{x}}}}})\cdot\delta\dot{\mathbf{x}}+\delta\dot{\mathbf{x}}\cdot\frac{\partial^{2}L}{\partial\dot{\mathbf{x}}\partial\dot{\mathbf{x}}}(\mathbf{\mathbf{\bar{\mathbf{x}}}},\dot{\mathbf{\mathbf{\bar{\mathbf{x}}}}})\cdot\delta\dot{\mathbf{x}}\right\} dt.\label{eq:functional Q}
\end{align}
In the following we will suppress the arguments of the derivatives
of the Lagranian. We will now recast Eq.~\eqref{eq:variation of OM}
in terms of self-adjoint operators using integration-by-parts and
$\delta\mathbf{x}(0)=\delta\mathbf{x}(T)=0$. We also note that $\left\langle \mathbf{f},P\frac{d}{dt}\mathbf{g}\right\rangle =-\left\langle \frac{d}{dt}\left(P^{T}\mathbf{f}\right),\mathbf{g}\right\rangle $,
for any matrix function $P(t)\in\mathbb{R}^{d\times d}$, and where
$\langle\mathbf{f},\mathbf{g}\rangle=\sum_{i}\int_{0}^{T}f_{i}(t)g_{i}(t)dt$,
which we use to symmetrise the second term in Eq.~\eqref{eq:functional Q}.
We get

\begin{equation}
S_{\text{OM}}[\bar{\mathbf{x}}+\delta\mathbf{x}]=S_{\text{OM}}[\bar{\mathbf{x}}(t)]+\langle\mathbf{j},\delta\mathbf{x}\rangle+\frac{1}{2}\langle\delta\mathbf{x},\mathcal{H}\delta\mathbf{x}\rangle+O(\delta\mathbf{x}^{3})\label{eq:OM expansion in operator form}
\end{equation}
where
\begin{align}
\mathbf{j}(t) & =\frac{\partial L}{\partial\mathbf{x}}-\frac{d}{dt}\frac{\partial L}{\partial\dot{\mathbf{x}}}\label{eq:euler-lagrange}\\
\mathcal{\mathcal{H}} & =-\frac{\beta}{2\mu}\frac{d^{2}}{dt^{2}}+2A(t)\frac{d}{dt}+B(t)\label{eq:Q operator}
\end{align}
and $A_{ij}(t)=\frac{\partial^{2}L}{\partial x_{[i}\partial\dot{x}_{j]}}$,
$B_{ij}(t)=\frac{\partial^{2}L}{\partial x_{i}\partial x_{j}}-\frac{d}{dt}\frac{\partial^{2}L}{\partial x_{j}\partial\dot{x}_{i}}$,
where closed brackets indicate an anti-symmetrisation over indices.
By completing the square, we find that Eq.~\eqref{eq:OM expansion in operator form}
defines a Gaussian process $\delta\mathbf{x}\sim\mathcal{N}(-\mathcal{H}^{-1}\mathbf{j},\mathcal{H}^{-1})$,
which describes the quadratic fluctuations around $\bar{\mathbf{x}}$.
In the sense of \citep{grossAbstractWienerSpaces1967}, the path-space
density of the Gaussian process is $\rho[\delta\mathbf{x}]\propto\exp(-\frac{1}{2}\langle\delta\mathbf{x}+\mathcal{H}^{-1}\mathbf{j},\mathcal{H}(\delta\mathbf{x}+\mathcal{H}^{-1}\mathbf{j})\rangle)$.
If the reference path solves the Euler-Lagrange equation Eq.~\eqref{eq:onsager-machlup action},
then $\mathbf{j}=0$ and the Gaussian process simplifies to $\rho[\delta\mathbf{x}]\propto\exp(-\frac{1}{2}\langle\delta\mathbf{x},\mathcal{\mathcal{\mathcal{H}}}\delta\mathbf{x})\rangle)$.

For systems with gradient dynamics $\mathbf{F}=-\nabla U$, the asymmetric
term in Eq.~\eqref{eq:Q operator} vanishes, and the form of the
operator simplifies to 
\begin{equation}
\mathcal{\mathcal{H}}=-\frac{\beta}{2\mu}\frac{d^{2}}{dt^{2}}+B(t)\label{eq:Q operator gradient}
\end{equation}

\subsection{Gaussian mixture approximation}

We now use the quadratic expansion of the Onsager-Machlup action to
construct an approximate probability measure over the TPE. \textit{\emph{Let
		$\mathbf{x}^{[\alpha]},\alpha=1,\dots,K$ be the local instantons
		of a given Langevin system. For each local instanton we can define
		a Gaussian measure $\mathbb{P}^{[\alpha]}$ with mean $\mathbf{x}^{[\alpha]}$
		and precision $H^{[\alpha]}$. Although the measure $\mathbb{P}^{[\alpha]}$
		is defined over the space of $C^{0}$ continuous paths, the distribution
		can be characterised via a density on the Hilbert space of $C^{2}$
		paths as $\rho^{[\alpha]}[\mathbf{x}]\propto\exp(-\langle\mathbf{x}-\mathbf{x}^{[\alpha]},\mathcal{H}^{[\alpha]}(\mathbf{x}-\mathbf{x}^{[\alpha]})\rangle)$
}}\citep{grossAbstractWienerSpaces1967,FunctionalIntegrationBasics}\textit{\emph{.
		To approximate the TPE distribution, we construct the mixed Gaussian
		density $\bar{\rho}[\mathbf{x}]\propto\sum_{\alpha=1}^{K}e^{-S_{\text{OM}}[\mathbf{x}^{[\alpha]}]}\rho^{[\alpha]}[\mathbf{x}]$.
		The normalisation constant $\mathcal{N}^{[\alpha]}$ of the densities
		$\rho^{[\alpha]}[\mathbf{x}]$ are not finite, but can be expressed
		as ratios with respect to the normalisation of the reference Wiener
		measure $\mathcal{N}^{W}$. This ratio can be shown to be equal to
		\citep{levitTheoremInfiniteProducts1977,dunneFunctionalDeterminantsQuantum2008}
		\begin{equation}
		\mathcal{Z}^{[\alpha]}:=\frac{\mathcal{N}^{[\alpha]}}{\mathcal{N}^{W}}=\left(\frac{\det[\mathcal{H}^{[\alpha]}]}{\det[-\frac{\beta}{2\mu}\frac{d^{2}}{dt^{2}}]}\right)^{-1/2}\label{eq:normalisation}
		\end{equation}
		where the RHS can be computed using the results in the Sec.}}~\textit{\emph{II-C.
		Using Eq.}}~\textit{\emph{\eqref{eq:normalisation} we can write
		down the approximation of the TPE as the Gaussian mixture \citep{gelmanBayesianDataAnalysis}
		\begin{equation}
		\mathbb{\bar{P}}=\sum_{\alpha=1}^{K}w_{\alpha}\mathbb{P}^{[\alpha]}\label{eq:TPE approximation}
		\end{equation}
		where $w_{\alpha}=e^{-S_{\text{OM}}[\mathbf{x}^{[\alpha]}]}\mathcal{Z}^{[\alpha]}/\sum_{\gamma=1}^{K}e^{-S_{\text{OM}}[\mathbf{x}^{[\gamma]}]}\mathcal{Z}^{[\gamma]}$.}}

\subsection{Approximations of transition channel probabilities}

We now derive an approximation to the probabilities of reactive pathways
using the Gaussian mixture approximation of the TPE. Let $\mathrm{E}_{\alpha}\subset C^{0},\alpha=1,\dots,K$,
which are disjoint open sets in the TPE, be the $K$ reactive pathways
under consideration. We define the observable
\begin{equation}
P^{[\alpha]}(\theta,T)=\mathbb{P}[E_{\alpha}]
\end{equation}
which is the probability of observing a path in $E_{\alpha}$. In
low temperatures we can assume that $\mathbb{P}(\cup_{\alpha}E_{\alpha})\approx1$,
i.e. approximately all stochastic paths transition via one of the
reactive pathways. Furthermore, we assume that $\mathbf{x}^{[\alpha]}\in E_{\alpha}$
and that $\mathbb{P}^{[\alpha]}(\cup_{\gamma\neq\alpha}E_{\gamma})\approx0$.
The latter assumption means that each measure $\mathbb{P}^{[\alpha]}$
is concentrated on $E_{\alpha}$ and lacks support on the other reactive
pathways. Under these assumptions we can approximate $P^{[\alpha]}(\theta,T)$
as
\begin{equation}
P^{[\alpha]}(\theta,T)\approx P_{G}^{[\alpha]}(\theta,T)\equiv w_{\alpha}.
\end{equation}

\subsection{Calculation of the Gaussian normalisation constants}

The regularised normalisation constants of Gaussians defined on functional
spaces can be found by computing the determinants of their covariance
operators. Equivalently, the normalisation can be found by computing
the determinant of their precision operator, which is the inverse
of the covariance operator. As for finite-dimensional linear operators,
determinants of differential operators can be found by computing their
eigenvalues, but this is in general a prohibitevely expensive computational
procedure. In the following we show that functional determinants,
acting on $d$-dimensional vectors, can be found by solving $d$ initial
value ODEs.

Let the linear operators
\begin{equation}
\mathcal{L}=\frac{d}{dt}\left(P\frac{d}{dt}\right)-R\label{eq:linear operator}
\end{equation}
and
\begin{equation}
\mathcal{L}_{0}=\frac{d}{dt}\left(P\frac{d}{dt}\right)\label{eq:free linear operator}
\end{equation}
be defined for $0\leq t\leq T$, and where $P(t)\in\mathbb{R}^{d\times d}$
is a positive-definite matrix function and $R(t)\in\mathbb{R}^{d\times d}$
is a matrix function. Let $\gamma^{(k)}$ and $\mathbf{u}^{(k)}(t;\alpha)$
be the eigenvalues and eigenfunctions of $\mathcal{L}$, which are
solutions to the boundary value problem
\begin{equation}
\mathcal{L}\mathbf{u}^{(k)}(t)=\gamma^{(k)}\mathbf{u}^{(k)}(t)\label{eq:eigenfunction eq}
\end{equation}
where $\mathbf{u}^{(k)}(0)=\mathbf{u}^{(k)}(T)=0$. Similarly, let
$\mathbf{u}_{0}^{(k)}(t)$ and $\gamma_{0}^{(k)}$ be the eigenfunctions
and eigenvalues of $\mathcal{L}_{0}$. Then the \emph{functional determinant}
of $\mathcal{L}$ is defined in regularised form as
\begin{equation}
\frac{\det\mathcal{L}}{\det\mathcal{L}_{0}}=\prod_{k=1}^{\infty}\frac{\gamma^{(k)}}{\gamma_{0}^{(k)}}.\label{eq:functional determinant}
\end{equation}
As the spectrum of Eq.~\eqref{eq:linear operator} is unknown, and
numerically expensive to compute, a much more efficient way of computing
Eq.~\eqref{eq:functional determinant} is via the Gelfand-Yaglom
theorem (GYT) \textit{\emph{\citep{gelfandIntegrationFunctionalSpaces1960,levitTheoremInfiniteProducts1977,dunneFunctionalDeterminantsQuantum2008}.
		The GYT states that the functional determinant can be expressed as
		\begin{equation}
		\left|\frac{\det\mathcal{L}}{\det\mathcal{L}_{0}}\right|=\left|\frac{\det\left[Y(T)\right]}{\det\left[Y_{0}(T)\right]}\right|\label{eq:GY result}
		\end{equation}
		where $Y(t)\in\mathbb{R}^{d\times x}$ with components $Y_{ij}(t)=y_{i}^{(j)}(t)$,
		where the $\mathbf{y}^{(j)}(t)$ are solutions to the $d$ second-order
		ODEs with initial conditions}}

\begin{align}
\mathcal{L}\mathbf{y}^{(j)}(t) & =0\label{eq:GY theorem}\\
\mathbf{y}^{(j)}(0) & =0\\
\frac{d}{dt}y_{i}^{(j)}(0) & =\delta_{ij}.
\end{align}
\textit{\emph{and where the matrix $Y_{0}(t)\in\mathbb{R}^{d\times d}$
		is defined similarly, but with $\mathcal{L}$ in Eq.}}~\textit{\emph{\eqref{eq:GY theorem}
		replaced with $\mathcal{L}_{0}$.}}

In the case of gradient dynamics, where $\mathcal{H}$ takes the form
in Eq.~\eqref{eq:Q operator gradient}, the GYT can be readily applied
by setting $P(t)=-\frac{\beta}{2\mu}I_{d}$ and $R(t)=-B(t)$, where
$I_{d}$ is the identity matrix. We now present a generalisation of
the GYT that allows for linear operators of the form

\begin{equation}
\mathcal{L}=\frac{d^{2}}{dt^{2}}+U\frac{d}{dt}+R\label{eq:linear operator with 1st order term}
\end{equation}
where $0\leq t\leq T$, and where $U(t),\ R(t)\in\mathbb{R}^{d\times d}$
are matrix functions, which then makes the GYT applicable for systems
non-gradient dynamics, with precision operators of the form Eq.~\eqref{eq:Q operator}.

We define the linear operator $\mathcal{G}$ which acts on vector
functions as
\[
\mathcal{G}\mathbf{y}=G\mathbf{y}
\]
where $G(t)$ is a matrix function that solves the equation

\begin{equation}
\dot{G}=-\frac{1}{2}UG,\label{eq:G eq}
\end{equation}
then the linear operator
\begin{equation}
\tilde{\mathcal{L}}=\mathcal{G}^{-1}\mathcal{L}\mathcal{G}=\frac{d}{dt^{2}}+G^{-1}\left(R-\frac{1}{2}\dot{U}-\frac{1}{4}U^{2}\right)G\label{eq:H tilde}
\end{equation}
is of the form Eq.~\eqref{eq:linear operator} and $\det\tilde{\mathcal{L}}$
can then be computed using the GYT. As for any two operators $\mathcal{A}$
and $\mathcal{B}$, we have that $\det\mathcal{A\mathcal{B}}=\det\mathcal{A}\det\mathcal{B}$,
and $\det\mathcal{A}^{-1}=1/\det\mathcal{A}$, and we therefore have
that $\det\mathcal{L}=\det\tilde{\mathcal{L}}$. The functional determinant
$\det\mathcal{L}$ can thus be computed by first solving Eq.~\eqref{eq:G eq},
constructing $\tilde{\mathcal{L}}$ using Eq.~\eqref{eq:H tilde},
and finally using the GYT to compute $\det\tilde{\mathcal{L}}$.

The above theorem can be applied to Eq. \eqref{eq:Q operator} by
setting $P(t)=-\frac{\beta}{2\mu}I_{d}$, $U=-\frac{4\mu}{\beta}A$
and $R=B$. Using the GYT, and its generalisation presented here,
the problem of computing the normalisation constants of Gaussian distributions
in functional spaces is reduced to solving $d$ initial value problems.

\section{MCMC method}

Here we define the MCMC method used to validate the semi-analytical
results on the transition path ensemble. We start with an introduction
to the \emph{preconditioned Crank-Nicholson} algorithm \citep{beskosMCMCMETHODSDIFFUSION2008,cotterMCMCMethodsFunctions2013,hairerSpectralGapsMetropolis2014},
and MCMC in continuous time. This is followed by a description of
the \emph{Teleporter MCMC}, which is the algorithm we use to sample
the systems introduced in the main text.

\subsection{MCMC in continuous time}

\begin{figure}[t]
	\includegraphics[width=1\textwidth]{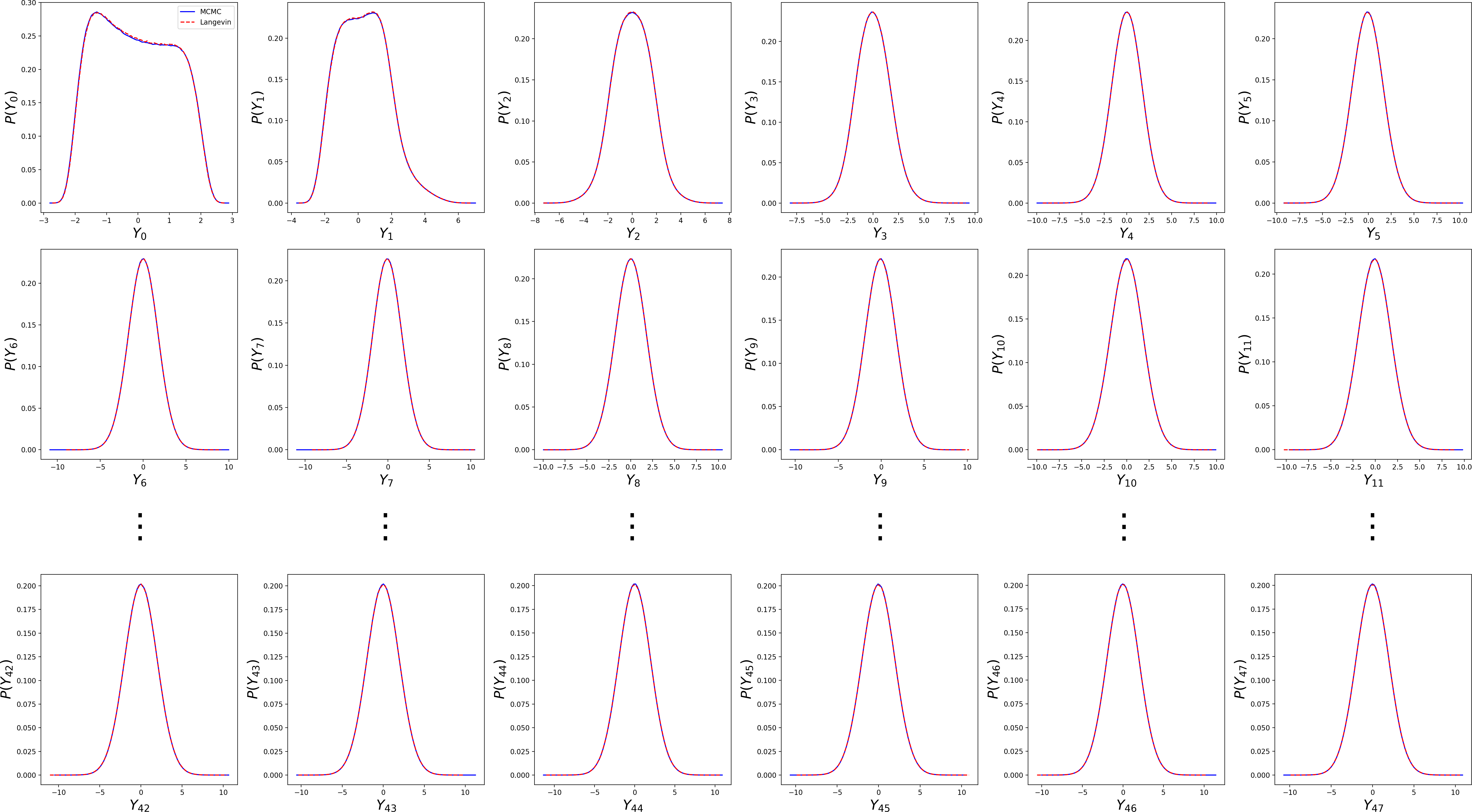} \centering
	\caption{The marginalised distributions of the of the modes of the sample paths
		of an asymmetric double well system, generated by the MCMC method
		(blue) and an Euler-Mayurama direct integrator (red) in the KKL basis,
		at $\theta/\theta_{0}=1.69$, $T/T_{0}=3.33$ and $N=200(T/T_{0})$.}
	\label{fig:MCMC vs langevin} 
\end{figure}

The Onsager-Machlup action Eq.~\eqref{eq:onsager-machlup action}
can be interpreted as a fictitious density $P[\mathbf{x}(t)]\propto\exp(-S_{\text{OM}}[\mathbf{x}(t)]$)
with respect to a fictituous Lebesgue measure on the space of continuous
paths \citep{takahashiProbabilityFunctionalsOnsagermachlup1981}\citep{grossAbstractWienerSpaces1967,FunctionalIntegrationBasics}.
This interpretation is however not mathematically rigorous: Neither
does an infinite-dimensional Lebesgue measure exist, nor is the Onsager-Machlup
action well defined for stochastic paths, which is because the temporal
integral over the term proportional to $\dot{x}^{2}$ diverges for
a stochastic path (which is typically nowhere differentiable). Both
these mathematical issues are fixed by absorbing the term proportional
to $\dot{x}^{2}$ into the measure, so that the path probability measure
$\mathbb{P}$ is described through its density \emph{with respect
	to} the Wiener measure $\mathbb{P}_{W}$. The resulting density is
given by the Girsanov formula $d\mathbb{P}/d\mathbb{P}_{W}=\exp(-\Phi[\mathbf{x}(t)])$,
where $\Phi$ is essentially the OM action minus the diverging $\dot{x}^{2}$
term, i.e.

\begin{align}
\Phi[\mathbf{x}(t)] & =\int_{0}^{T}(\frac{\beta\mu}{4}\mathbf{F}^{2}+\frac{\mu}{2}\nabla\cdot\mathbf{F})dt-\frac{\beta}{2}\int_{\mathbf{x}_{0}}^{\mathbf{x}_{T}}\mathbf{F}\cdot d\mathbf{X}.\label{eq:girsanov}
\end{align}
This is a functional on stochastic paths in the TPE \citep{itoProbabilisticConstructionLagrangean1978a,pavliotisStochasticProcessesApplications2014,gladrowExperimentalMeasurementRelative2021}.
The second term in Eq.~\eqref{eq:girsanov} is interpreted as an
Itô integral. Both integrals in Eq.~\eqref{eq:girsanov} are well-defined
and finite when evaluated on stochastic paths. 

As the Wiener measure $\mathbb{P}_{W}$ and the Girsanov formula $d\mathbb{P}/d\mathbb{P}_{W}=\exp(-\Phi[\mathbf{x}(t)])$
provide a mathematically well-defined description of the probability
distribution on path space induced by Langevin dynamics, they form
a natural starting point for path-sampling algorithms for the TPE.
The strategy we follow here is to obtain samples in the target measure
$\mathbb{P}$ of the Itô process, by evaluating its density $\exp(-\Phi)$
relative to the Gaussian measure $\mathbb{P}_{W}$ \citep{beskosMCMCMETHODSDIFFUSION2008,hairerSpectralGapsMetropolis2014,cotterMCMCMethodsFunctions2013}.
This is implemented through the pre-conditioned Crank-Nicholson (pCN)
Markov Chain Monte Carlo (MCMC) procedure as follows \citep{beskosMCMCMETHODSDIFFUSION2008,cotterMCMCMethodsFunctions2013,hairerSpectralGapsMetropolis2014}.
Abstractly, at the $n$-th iteration a sample $\mathbf{W}^{(n)}$
is drawn from the reference Gaussian measure $\mathbb{P}_{W}$, and
a Metropolis-Hastings proposal $\tilde{{\mathbf{X}}}^{(n+1)}=\sqrt{1-\kappa^{2}}\mathbf{X}^{(n)}+\kappa\mathbf{W}^{(n)}$
(where $0<\kappa\leq1$) is constructed from the currentsample transition
path $\mathbf{X}^{(n)}$. With probability $p_{a}=\text{min\ensuremath{\left\{  1,\exp(\Phi[\mathbf{X}^{(n)}]-\Phi[\tilde{{\mathbf{X}}}^{(n+1)}])\right\} } }$
this proposal is accepted, ${\mathbf{X}}^{(n+1)}=\tilde{{\mathbf{X}}}^{(n+1)}$.
If the proposal is rejected the current transition path is retained,
${\mathbf{X}}^{(n+1)}={\mathbf{X}}^{(n)}$. 

Concretely, to sample from the reference Gaussian measure $\mathbb{P}_{W}$,
we use its Kosambi-Karhunen-Loève (KKL) expansion \citep{kosambiParallelismPathspaces2016,karhunenUberLineareMethoden1947,loeveProbabilityTheory1977}

\begin{equation}
{\bf W}^{(n)}(t)=\mathbf{x}_{0}+\bar{\mathbf{v}}t+\sqrt{\frac{2\mu}{\beta}}\sum_{i=1}^{\infty}\mathbf{Z}_{i}^{(n)}\sqrt{\lambda_{i}}\phi_{i}(t)\label{eq:kkl expansion}
\end{equation}
which starts at $\mathbf{x}_{0}$ and ends at $\mathbf{x}_{T}$, where
$\lambda_{i}=T^{2}/\pi^{2}i^{2}$, $\phi_{i}(t)=\sqrt{2/T}\sin(t/\sqrt{\lambda_{i}})$,
$\bar{\mathbf{v}}=(\mathbf{x}_{T}-\mathbf{x}_{0})/T$ and $\mathbf{Z}_{i}^{(n)}$
are independent and identically distributed zero-mean unit-variance
normal random variables. With the expansion Eq.~\eqref{eq:kkl expansion}
generating a sample Gaussian process means drawing a sample of coefficients
$\mathbf{Z}_{i}^{(n)}$. In a key step, which is a new contribution
in this Letter, we also expand the stochastic path ${\bf X}^{(n)}(t)$
in the same basis as
\begin{equation}
{\bf X}^{(n)}(t)=\mathbf{x}_{0}+\bar{\mathbf{v}}t+\sqrt{\frac{2\mu}{\beta}}\sum_{i=1}^{N}\mathbf{Y}_{i}^{(n)}\sqrt{\lambda_{i}}\phi_{i}(t).\label{eq:path expansion}
\end{equation}
where $N$ is a truncation of the expansion to render the algorithm
amenable for numerics. Because the Itô process contains a drift term,
Eq.~\eqref{eq:path expansion} is not a KKL expansion of the Itô
process, but a parametrisation of a stochastic path in the TPE in
a countably infinite basis and continuous time. The $\mathbf{Y}_{i}^{(n)}$
are in general not zero-mean unit-variance normal random variables,
and their distributions must be obtained through the MCMC procedure.
The linearity of the Metropolis-Hastings pCN proposal and the orthogonality
of the basis functions $\phi_{i}(t)$ implies that proposals can be
defined directly in coefficent-space $\mathbf{Y}_{i}^{(n+1)}=\sqrt{1-\kappa^{2}}\mathbf{Y}_{i}^{(n)}+\kappa\mathbf{Z}_{i}^{(n)}$,
with the acceptance probability being determined as before. We also
note that the KKL basis allows us to exploit the fast Fourier transform
to evaluate Eq.~\eqref{eq:path expansion} giving an $O(N\log N)$
efficiency for $N$ degrees of freeedom. 

To summarize, a step of the pCN algorithm consists of i) sampling
from the Wiener measure by drawing a multivariate Gaussian random
variable $\mathbf{Z}_{i}^{(n)}$, ii) calculating the corresponding
proposed path coefficients $\mathbf{Y}_{i}^{(n+1)}$ (which define
a proposed transition path via Eq.~\eqref{eq:path expansion}), iii)
evaluating the Girsanov functional on the proposed path $\Phi[\mathbf{\tilde{X}}^{(n+1)}(t)]$,
and iv) accepting the proposed path with probability $p_{a}$.

\subsection{Validation of MCMC method}

We now compare the results of the MCMC method against a direct Euler-Mayurama
integrator on an example system, to verify the validity of the former.
As direct methods are generally very inefficient for sampling TPEs,
we performed the numerical verification on a simple one-dimensional
system, rather than the two-dimensional systems in the main text.
Fig. \ref{fig:MCMC vs langevin} shows a comparison of the results
of the MCMC method against a direct simulation of the Langevin equation
of an asymmetric double-well potential of the form

\begin{equation}
U(x)=U_{0}\left(\left(\frac{x}{L}-1\right)^{2}-\frac{1}{4}\frac{\Delta U}{U_{0}}\left(\frac{x}{L}-2\right)\right)\left(\frac{x}{L}+1\right)^{2}\label{eq:asymmetric double well}
\end{equation}
where we have set $\frac{\Delta U}{U_{0}}=1/2$. We generated samples
of the transition path ensemble with $\tilde{x}(0)=-1$ and $\tilde{x}(T)=1$.
For the Euler-Mayurama method we generated samples by collecting trajectories
with end-points within a small interval $\tilde{x}(T)\in[1-\epsilon,1+\epsilon]$
around the right minima, where $\epsilon=10^{-2}$. As the results
of Fig. \ref{fig:MCMC vs langevin} demonstrates the validity of the
MCMC method, we can use the latter to verify the semi-analytical Gaussian
mixture approximations.

\subsection{Teleporter MCMC}

\begin{figure}[t]
	\includegraphics[width=0.66\textwidth]{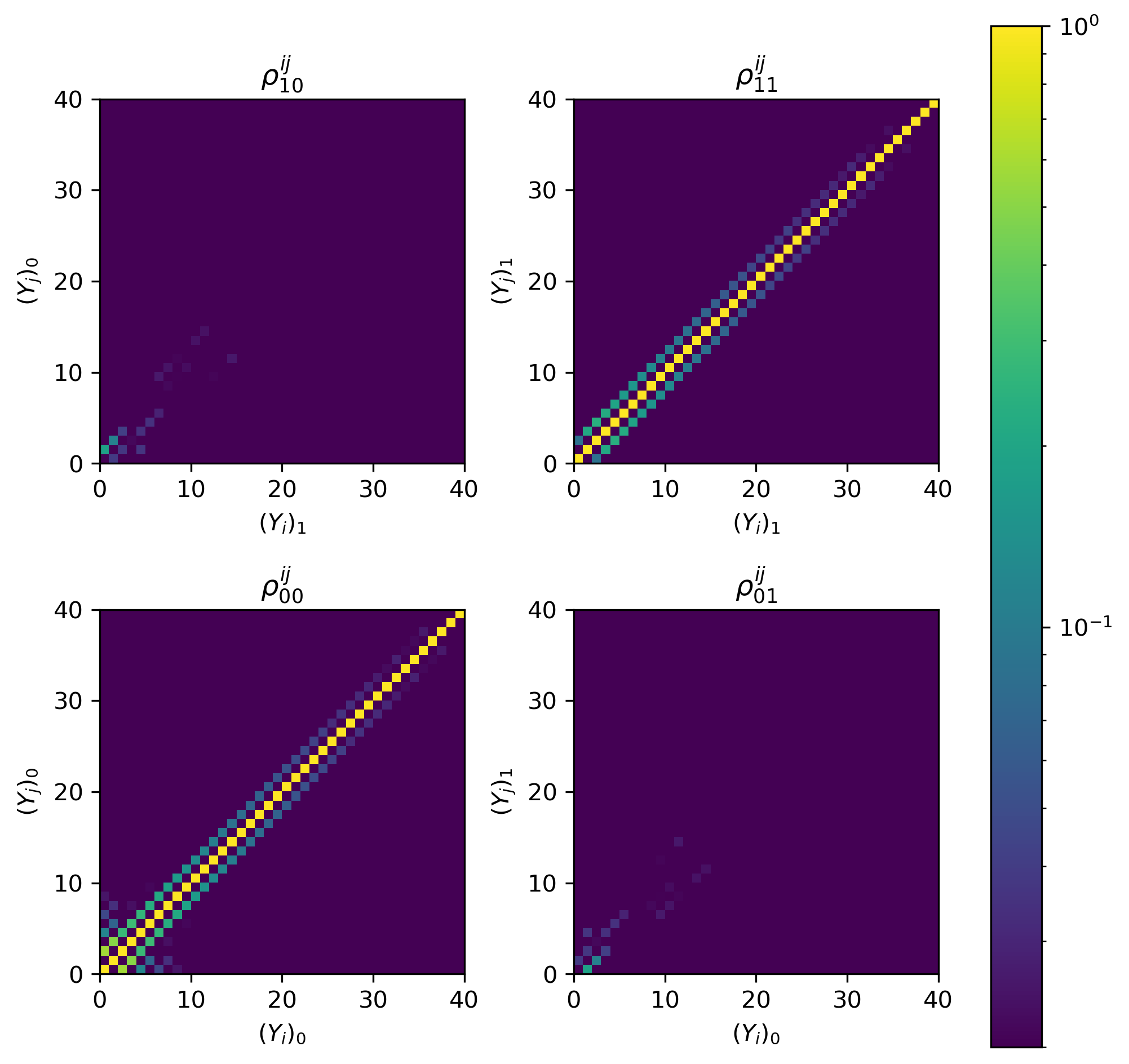}
	\centering \caption{The absolute normalised covariance $\rho_{kl}^{ij}=E^{\mathbb{P}}\left\{ (Y_{i})_{k}(Y_{j})_{l}\right\} /\sqrt{E^{\mathbb{P}}\left\{ (Y_{i})_{k}^{2}\right\} E^{\mathbb{P}}\left\{ (Y_{j})_{l}^{2}\right\} }$
		of the modes of the sample paths in the KKL basis, found by sampling
		the system with gradient dynamics using the TMC, at $\theta/\theta_{0}=3.36$,
		$T/T_{0}=3$ and $N=200(T/T_{0})$.}
	\label{fig:mode covariance} 
\end{figure}
\begin{figure}[t]
	\includegraphics[width=1\textwidth]{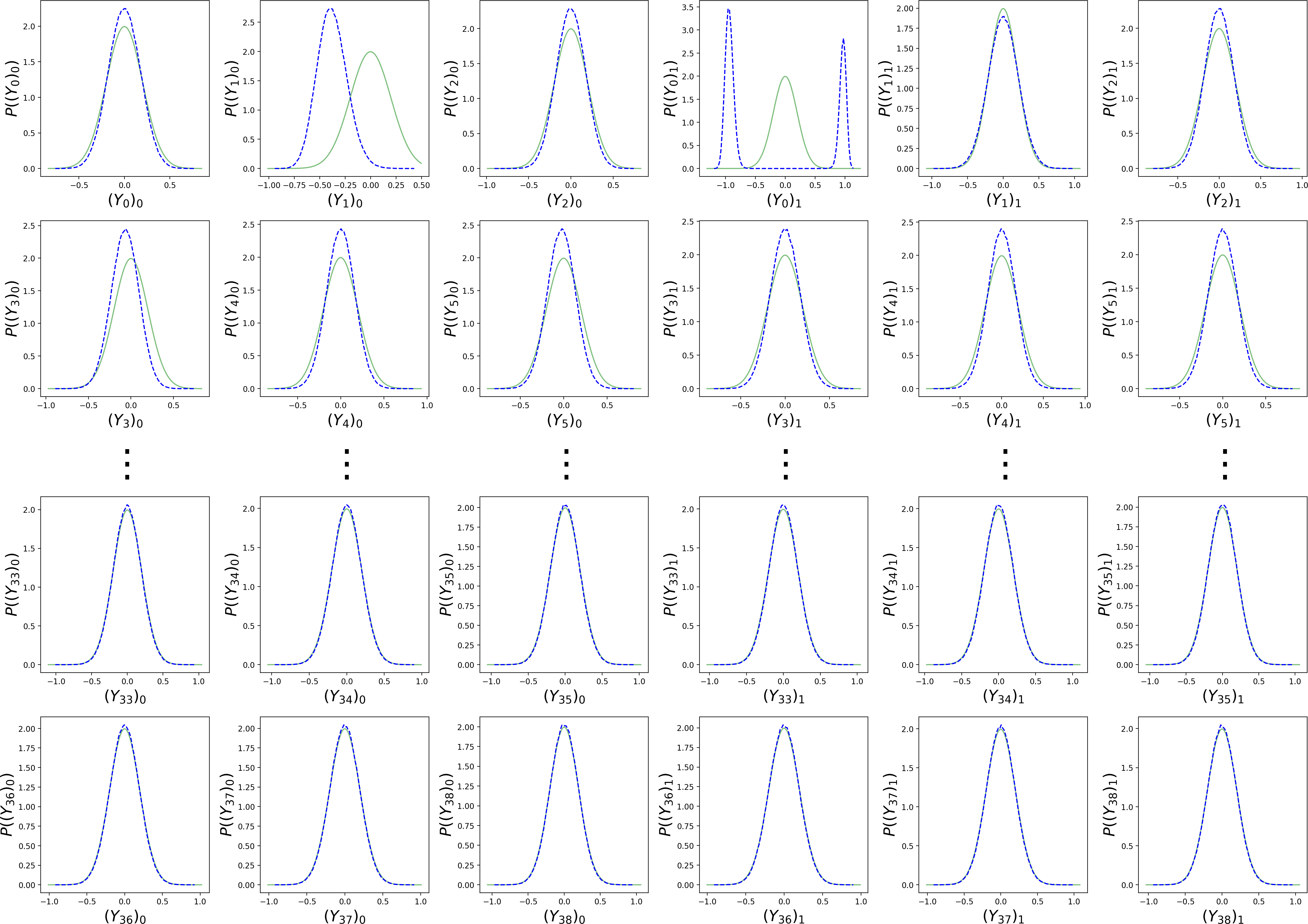}
	\centering \caption{The marginalised distributions of the of the modes of the sample paths
		(blue) and the Wiener process (green) in the KKL basis, found by sampling
		the system with gradient dynamics using the TMC, at $\theta/\theta_{0}=3.36$,
		$T/T_{0}=3$ and $N=200(T/T_{0})$.}
	\label{fig:mode marginalised} 
\end{figure}
Here we describe in further detail the \emph{Teleporter MCMC }(TMC)
algorithm used in the main text. We start with a discussion of the
algorithm in its infinite dimensional form, defined directly on the
space of continuous paths, and then proceed to describe a modification
of the algorithm adapted to the KKL discretisation.

At each step of the MCMC, with a probability $p_{\text{teleport}}$,
we draw an independent proposal step $\mathbf{X}^{'}$ from the mixed
Gaussian distribution

\begin{equation}
\bar{\mathbb{P}}=\sum_{\alpha=1}^{K}w_{\alpha}\mathbb{\mathbb{P}}^{[\alpha]}\label{eq:mixed gaussian-1}
\end{equation}
where the weights $w_{\alpha}$ are parameters that must satisfy $\sum_{\alpha=1}^{K}w_{\alpha}=1$,
and where $\mathbb{\mathbb{P}}^{[\alpha]}$ are Gaussian distributions
with precision operators $\mathcal{H}^{[\alpha]}$ and mean $\bar{\mathbf{x}}$
as defined in previous sections. Using the Metropolis-Hastings condition
to ensure that the MCMC samples the target measure $\mathbb{P}$ of
the Itô process we find that $\mathbf{X}'$ should be accepted with
probability
\begin{equation}
a_{\text{TMC}}\left[\mathbf{X}',\mathbf{X}^{(n)}\right]=\min\left\{ 1,\exp\left(\Phi[\mathbf{X}^{(n)}]-\Phi[\mathbf{X}']\right)\frac{\sum_{\alpha}w_{\alpha}\exp\left(-\Psi^{[\alpha]}[\mathbf{X}^{(n)}]\right)}{\sum_{\alpha}w_{\alpha}\exp\left(-\Psi^{[\alpha]}[\mathbf{X}']\right)}\right\} \label{eq:acceptance probability}
\end{equation}
where $\mathbf{X}^{(n)}$ is current state of the MCMC, and $\Psi^{[\alpha]}$
is the logarithmic density of $\mathbb{P}^{[\alpha]}$ with respect
to the Wiener measure $\frac{d\mathbb{\mathbb{P}}^{[\alpha]}}{d\mathbb{P}_{W}}=\exp(-\Psi^{[\alpha]})$
and

\begin{align}
\Psi^{[\alpha]}[\mathbf{X}] & =\Psi_{1}^{[\alpha]}[\mathbf{X}-\mathbf{x}^{[\alpha]}]+\Psi_{2}^{[\alpha]}[\mathbf{X}]\\
\Psi_{1}^{[\alpha]}[\mathbf{X}] & =\int_{0}^{T}\left(2\mathbf{X}^{T}A(t)d\mathbf{X}+\mathbf{X}^{T}B(t)\mathbf{X}dt\right)\\
\Psi_{2}^{[\alpha]}[\mathbf{X}] & =\frac{\beta}{2\mu}\int_{0}^{T}\left(2\dot{\mathbf{x}}^{[\alpha]T}d\mathbf{X}-|\dot{\mathbf{x}}^{[\alpha]}|^{2}dt\right)
\end{align}
Thus far, the algorithm has been defined directly on the space of
continuous paths, but in numerical applications it is necessary to
apply a discretisation procedure. We can approximate $\mathbb{\mathbb{P}}^{[\alpha]}$
as a multivariate Gaussian by expanding its precision operator $\mathcal{H}^{[\alpha]}$
in the KKL basis \citep{kosambiParallelismPathspaces2016,karhunenUberLineareMethoden1947,loeveProbabilityTheory1977}
of the Wiener process as $\left(H_{ij}^{[\alpha]}\right)_{kl}=\langle\mathbf{e}_{k}\phi_{i},$$\mathcal{H}^{[\alpha]}\mathbf{e}_{l}\phi_{j}\rangle$,
$i,j=1,\dots,N$, $k,l=1,\dots,d$, where, $\mathbf{e}_{k}$ is a
constant vector with one non-zero component $(e_{k})_{l}=\delta_{kl}$.
Due to our discretisation procedure $M$ can be kept small as the
noise dominates over the drift for high-frequency modes, which manifests
itself as $H_{ij}^{[\alpha]}$ rapdily converging onto the precision
matrix of the Wiener measure , $\left(H_{ij}^{W}\right)_{kl}=\frac{\beta}{4\mu}\delta_{ij}\delta_{kl}$,
for high mode numbers $i,j$. This is demonstrated in Fig. \ref{fig:mode covariance}
and Fig. \ref{fig:mode marginalised}.

Using the multivariate Gaussian with precision matrix $\tilde{H}^{[\alpha]}$,
we construct a grafted Gaussian process ${\bf W}^{[\alpha]}(t)=\mathbf{x}^{[\alpha]}(t)+\sqrt{\frac{2\mu}{\beta}}\sum_{i=1}^{\infty}\mathbf{Z}_{i}^{[\alpha]}\sqrt{\lambda_{i}}\phi_{i}(t)$
where $(\mathbf{Z}_{1}^{[\alpha]},\dots,\mathbf{Z}_{M}^{[\alpha]})\sim\mathcal{N}(0,\tilde{H}^{[\alpha]})$
and $\mathbf{Z}_{i}^{[\alpha]}\sim\mathcal{N}(0,I_{d}),\ i>M$, where
$I_{d}$ is the $d$-dimensional identity matrix. This defines a Gaussian
measure $\mathbb{\tilde{P}}^{[\alpha]}$ on the space of stochasic
paths from which we can sample efficiently. Finally, we construct
a Gaussian mixture measure as the linear combination

\begin{equation}
\tilde{\mathbb{P}}=\sum_{\alpha=1}^{K}\tilde{w}_{\alpha}\mathbb{\tilde{P}}^{[\alpha]}\label{eq:mixed gaussian-1-1}
\end{equation}
from which we draw independent samples in the same manner as before.
The logarithmic densities of $\mathbb{\tilde{P}}^{[\alpha]}$ with
respect to the Wiener measure $\mathbb{P}_{W}$ are now

\begin{equation}
\tilde{\Psi}^{[\alpha]}[\mathbf{X}]=\sum_{i,j=1}^{M}\sum_{k,l=1}^{d}\frac{1}{2}(Y_{ik}-Y_{ik}^{[\alpha]})\tilde{K}_{ijkl}^{[\alpha]}(Y_{jl}-Y_{jl}^{[\alpha]})+2\sum_{i=1}^{M}\mathbf{Y}_{i}^{[\alpha]T}\mathbf{Y}_{i}+\sum_{i=1}^{M}\mathbf{Y}_{i}^{[\alpha]T}\mathbf{Y}_{i}^{[\alpha]}.
\end{equation}
where ${\bf X}(t)=\mathbf{x}_{0}+\bar{\mathbf{v}}t+\sqrt{\frac{2\mu}{\beta}}\sum_{i=1}^{\infty}\mathbf{Y}_{i}\sqrt{\lambda_{i}}\phi_{i}(t)$,
$Y_{ik}$ denotes the $k$th component of $\mathbf{Y}_{i,}$ $Y_{ik}^{[\alpha]}=\langle\mathbf{x}^{[\alpha]},\mathbf{e}_{k}\phi_{i}\rangle$,
$\left(\tilde{K}_{ij}^{[\alpha]}\right)_{kl}=\langle\mathbf{e}_{k}\phi_{i},(\mathcal{H}^{[\alpha]}-\mathcal{H}^{W})\mathbf{e}_{l}\phi_{j}\rangle$,
$\tilde{L}_{ik}^{[\alpha]}=\frac{\beta}{2\mu}$ using which the acceptance
probabilities are computed as in Eq.~\eqref{eq:acceptance probability}.

We now summarise the full algorithm, expressed in the KKL basis:

\begin{enumerate}
	
	\item Choose an initial state $\mathbf{Y}^{(0)} \in \mathbb{R}^{N \times d}$.
	
	\item Draw a random number $U^{(n+1)} \sim \text{Unif}([0,1])$.
	
	\begin{itemize}
		
		\item If $U^{(i+1)} > p_\text{teleport}$.
		\begin{enumerate}
			
			\item Given state $\mathbf{Y}^{(n)}$, the $(n+1)$-th proposal is
			\begin{equation}
			\mathbf{Y}'_i = \sqrt{1 - \kappa^2} \mathbf{Y}_i^{(n)} + \kappa \mathbf{Z}_i^{(n)}
			\end{equation}
			where $\mathbf{Z}_i^{(n)} \sim \mathcal{N}(0,I_d)$ and $i=1,\dots,N$.
			
			\item Draw a random number $V^{(n+1)} \sim \text{Unif}([0,1])$.
			\begin{itemize}
				\item If $V^{(i+1)} < a[\mathbf{x}(t;\mathbf{Y}'), \mathbf{x}(t;\mathbf{Y}^{(n)})]$ then set $\mathbf{Y}^{(n+1)} = \mathbf{Y}'$.
				\item Otherwise set  $\mathbf{Y}^{(n+1)} = \mathbf{Y}^{(n)}$.
			\end{itemize}
			
		\end{enumerate}
		
		\item If $U^{(i+1)} \leq p_\text{teleport}$:
		
		\begin{enumerate}
			\item Given state $\mathbf{Y}^{(n)}$, the $(n+1)$-th proposal is
			\begin{equation}
			\mathbf{Y}'_i = \tilde{\mathbf{Z}}_{i}^{(n)}
			\end{equation}
			where $\tilde{\mathbf{Z}}^{(n)}$ is drawn from $\tilde{\mathbb{P}}_N$, and  $i=1,\dots,N$.
			
			\item Draw a random number $W^{(n+1)} \sim \text{Unif}([0,1])$.
			\begin{itemize}
				\item If $W^{(i+1)} < a_\text{TMC}[\mathbf{x}(t;\mathbf{Y}'), \mathbf{x}(t;\mathbf{Y}^{(n)})]$ then set $\mathbf{Y}^{(n+1)} = \mathbf{Y}'$.
				\item Otherwise set  $\mathbf{Y}^{(n+1)} = \mathbf{Y}^{(n)}$.
			\end{itemize}
		\end{enumerate}
		
	\end{itemize}
	
	\item Repeat step 2.
	
\end{enumerate}where $\text{Unif}([0,1])$ is the uniform distribution over the unit
interval, $\mathbf{x}(t;\mathbf{Y})=\mathbf{x}_{0}+\bar{\mathbf{v}}t+\sqrt{\frac{2\mu}{\beta}}\sum_{i=1}^{N}\mathbf{Y}_{i}\sqrt{\lambda_{i}}\phi_{i}(t)$,
and $\tilde{\mathbb{P}}_{N}$ is the truncation of Eq.~\eqref{eq:mixed gaussian-1-1}
to $N$ modes. In the numerical experiments discussed in the main
text, we used $\tilde{w}_{1}=\tilde{w}_{2}=1/2$ and $M=10(T/T_{0})$.

As mentioned in the main text, an alternate method to the above would
be a synthesis of the two sampling approaches in the above algorithm.
We could replace the reference Wiener measure $\mathbb{P}_{W}$ with
the mixed Gaussian $\tilde{\mathbb{P}}$, and thus perform the pCN-MCMC
with $\tilde{\mathbb{P}}$ as the invariant measure.

\end{document}